\documentclass[sn-mathphys,Numbered]{sn-jnl} 
\usepackage{graphicx}%
\usepackage{multirow}%
\usepackage{amsmath,amssymb,amsfonts}%
\usepackage{amsthm}%
\usepackage{mathrsfs}%
\usepackage[title]{appendix}%
\usepackage{xcolor}%
\usepackage{textcomp}%
\usepackage{manyfoot}%
\usepackage{booktabs}%
\usepackage{algorithm}%
\usepackage{algorithmicx}%
\usepackage{algpseudocode}%
\usepackage{listings}%
\usepackage{bm}
\usepackage{hyperref}

\unnumbered

\begin{document}

\title[Article Title]{Exotic swarming dynamics of high-dimensional swarmalators}

\author[1]{\fnm{Akash} \sur{Yadav}}\email{akashyadav23@iisertvm.ac.in}

\equalcont{These authors contributed equally to this work.}

\author[1]{\fnm{Krishnanand} \sur{J}}\email{krishnanandj2000@gmail.com}
\equalcont{These authors contributed equally to this work.}

\author[2]{\fnm{V.K.} \sur{Chandrasekar}}\email{chandru25nld@gmail.com}
\author[3,4]{\fnm{Wei} \sur{Zou}}\email{weizou83@gmail.com}
\author[5,6]{\fnm{J\"{u}rgen } \sur{Kurths}}\email{kurths@pik-potsdam.de}
\author*[1]{\fnm{D.V.} \sur{Senthilkumar}}\email{skumar@iisertvm.ac.in}

\affil*[1]{\orgdiv{School of Physics}, \orgname{Indian Institute of Science Education and Research Thiruvananthapuram}, \orgaddress{ \city{Thiruvananthapuram}, \postcode{695551}, \state{Kerala}, \country{India}}}

\affil[2]{\orgdiv{Department of Physics, Center for Nonlinear Science and Engineering}, \orgname{School of Electrical and Electronics Engineering, SASTRA Deemed University}, \orgaddress{ \city{Thanjavur}, \postcode{613401}, \state{Tamil Nadu}, \country{India}}}

\affil[3]{\orgdiv{School of Mathematical Sciences}, \orgname{South China Normal University}, \orgaddress{\city{Guangzhou} \postcode{510631},\country{ China}}}

\affil[4]{\orgdiv{Research Institute of Intelligent Complex Systems}, \orgname{Fudan University}, \orgaddress{\city{Shanghai}, \postcode{200433},\country{ China}}}

\affil[5]{\orgname{Potsdam Institute for Climate Impact Research}, \orgaddress{\city{Telegraphenberg Potsdam }, \postcode{D-14415},\country{ Germany}}}

\affil[6]{\orgdiv{Institute of Physics}, \orgname{Humboldt University Berlin}, \orgaddress{\city{Berlin}, \postcode{D-12489}, \country{ Germany}}}

\abstract{Swarmalators are oscillators that can swarm as well as sync via a dynamic balance between their spatial proximity and phase similarity.
We present a generalized D-dimensional swarmalator model, which is more realistic and versatile, that captures self-organizing behaviors
of a plethora of real-world collectives. This allows for modeling complicated processes such as flocking, schooling of fish, cell sorting during embryonic development,residential segregation, and opinion dynamics in social groups. We demonstrate its versatility by capturing the manoeuvers of school of fish and traveling waves of gene expression, both qualitatively and quantitatively,  embryonic cell sorting,  microrobot collectives and various life stages of slime mold by a suitable extension of the original model to incorporate appropriate features  besides a gallery of its intrinsic self-organizations for various interactions.  
We expect this high-dimensional  model to be potentially useful in describing swarming systems in a wide range of disciplines
including physics of active matter, developmental biology, sociology, and engineering.}

\keywords{Swarmalators, Complex Systems, Synchronization, Self-Organization}

\maketitle
\section{Introduction}\label{sec1}

Complex systems science aims at unravelling  the underlying dynamical processes  that are responsible for a plethora of self-organizing collective 
behaviors in various branches of science and technology  including systems biology~\cite{jrap2009,MUMA2018}, climate science~\cite{FJMJ2020,bnr2019}, 
complex networks~\cite{PSRC2015,Liu2011}, ecology~\cite{VGER2005} and social studies~\cite{hscp2021,vsas2016} in an effort towards
the effective utilization of the natural resources,  and their sustainability. A recent strongly emerging interest in complex systems science, which is gaining momentum, 
are studies on the swarming dynamics~\cite{kpokhh2017,gkssnc2022,sckok2023,fjm2020,pjfnm2022,nbal2022,sykpok2022,gksdg2023}.  
Swarmalators represent a class of systems which are able to self-aggregate spatially  (swarm) and simultaneously
adjust their internal rhythms (synchronize) through a delicate balance between their spatial proximity and phase similarity, representing the latter. Pioneering 
contributions were  recently made by Igoshin et al. and Tanaka et al.  in modeling the dynamics of chemotactic oscillators~\cite{oaiam2001,dt2007,miki2012} 
and by Levis et al. in describing the dynamics of revolving agents~\cite{dlbl2019,dlip2019}. 

Recent surge in the studies on the swarmalators displayed a zoo of collective dynamical states mimicking the self-organizing dynamics of 
natural~\cite{ihrkk2005,jmsos2013} and technological~\cite{ys2012,kh2020,jyscb2015} collectives ranging from  spermatoza~\cite{ihrkk2005} to drones and 
robots~\cite{abcb2020,mzm2020,mst2021}.  
A recent study by O\textquotesingle Keeffe et al.~\cite{kpokhh2017} elucidated that static phase wave (SPW)  is qualitatively similar to the `asters' formed by the ferromagnetic colloids~\cite{asisa2011}, 
whereas the active phase wave (APW)  has the characteristic features of `vortex arrays' formed by the populations of spermatoza~\cite{ihrkk2005}.  
Generalization of the swarmalator model in~\cite{kpokhh2017}  by including  non-identical frequencies, chirality  and local coupling,  unveiled several new 
spatiotemporal regimes including  interacting phase waves, vortices, and beating clusters~\cite{sckok2023}. It was also shown
that many of these self-organizing patterns qualitatively resemble those exhibited by cellular self-organization~\cite{cdtaa2016,kulgm2017},
flocking patterns of Quinke rollers~\cite{kh2020,bzas2020},  and the various life stage of slime mold~\cite{ls2006}.
An attempt for an analytical description of the
synchronized state and the existence condition for a few of the states using a basic model was presented recently~\cite{sykpok2022}.  
Other extensions by including Gaussian function for  short-range repulsive interaction~\cite{fjm2020}, delayed interactions~\cite{nbal2022}, and  pinning~\cite{gksdg2023}
have also been made. 

However, immediate applications of the observed swarming dynamics remain elusive despite the majority of them qualitatively  resemble  the emerging 
patterns of a wide variety of real-world systems. Nevertheless, a direct comparison of several measurable observables from the minimalistic swarmalator model to that of 
experimental data elucidate that real-world applications of swarmalator models are imminent.  
One of the prospective  applications of a reconfigurable swarmalator dynamics is their locomotive utility~\cite{kpokhh2017}. 
For instance, the collective metachronal waves known to
facilitate biological transport~\cite{lwifm1993,jegg2013} of the populations of cilia are similar to APW and splintered phase wave (SPPW).  
Particularly, reconfigurable microrobot swarms,  a perfect real-world example of swarmalators,  can  be used for biomedical applications including targeted 
drug delivery~\cite{mzm2020,mst2021,ggsc2022}.  

Given the tremendous potential real-world applications of swarmalators, the model employed in the aforementioned studies invariably comprises of evolution equations
for two-dimensional spatial variables and  one-dimensional Kuramoto model~\cite{jaallb2005} governing the phase dynamics.  However, most of these studies generalized their results to three-dimensional spatial variables, where the phase evolves  on a unit circle in accordance with  the spatial proximity but lacks  the orientational degree of freedom. But this is a key component of almost all real-world swarmalators such as  spin orientation in ferromagnetic collides~\cite{akas2017}, velocity vector of birds flock~\cite{tvac1995,csg2019}, fish schools,  and swarm of drones. The orientation degree of freedom is an intrinsic feature of all systems described by spherical polar coordinates, in which the orientation vector is specified by both polar angle $\theta$ and azimuthal angle $\phi$. The internal state of such systems is inevitably described by the orientation vector represented in terms of $\theta$ and $\phi$  in three-dimensions (3D), which is missing in the existing studies on swarmalators. 

To treat this important point, we introduce here a D-dimensional swarmalator model governed by D-dimensional spatial and orientation vectors, in general, for predictive fidelity  
of the self-organizing behaviors of real-world swarmalators, where the alignment of their orientation vectors represents their intrinsic dynamics. 
For more sensible visualization  and interpretation of results, we restrict our simulations to 3D space and 3D  phase variables.
We show that in view of the  inseparable dynamics of $\theta$ and $\phi$  affecting the spatial proximity  and vice versa in 3D, our model  indeed facilitates a 
repertoire of exotic self-organizing behaviors (see Table S1) that are specific to our model in addition to those states observed by the aforementioned studies with similar settings.

\section{Results}\label{sec2}
\subsection*{The Model}
The proposed D-dimensional  swarmalator model is represented by
\begin{subequations}
\begin{equation}
    \bm{\dot{x}}_i = \bm{v}_i + \frac{1}{N-1}\sum_{j=1} ^N \left [ \{1+J (\bm{\sigma}_i \cdot \bm{\sigma}_j)\} \frac{\bm{x}_{j}-\bm{x}_{i}}{\left | \bm{x}_{j} -\bm{x}_{i} \right |^\alpha} - \frac{\bm{x}_{j}-\bm{x}_{i}}{\left | \bm{x}_{j}-\bm{x}_{i} \right |^\beta}  \right ]+\bm{\xi}^{\bf{x}}_i(t), \label{space}
\end{equation}
\begin{equation}
    \bm{\dot{\sigma}}_i = \bm{W}_i  \bm{\sigma}_i + \sum_{j=1}^N K_{ij} \left[ \frac{\bm{\sigma}_j-(\bm{\sigma}_j\cdot \bm{\sigma}_i) \bm{\sigma}_i}{\left | \bm{x}_{j}-\bm{x}_{i} \right |^\gamma} \right]+\bm{\xi}^{\bm{\sigma}}_i(t), \label{phase}
\end{equation}
\end{subequations}
where $ i = 1,2,3,\ldots, N$ is the number of swarmalators, $\bm{x}_i$  is the  D-dimensional position vector of the $i^{th}$swarmalator, $\bm{\sigma}_i$ is its orientation vector on the D-dimensional unit hyper-sphere characterizing the intrinsic dynamics of the swarms, and $\bm{v_i}$ is its  self-propulsion velocity.  
Note that the evolution equation for $\bm{\sigma}_i$ in the absence of the distant dependent kernel is the D-dimensional Kuramoto model~\cite{csg2019,kkar2021}
(see supplementary text S1 for its derivation).  
 In the context of flocking and swarming models $\bm{\sigma}_i$ can be interpreted  as the unit vector along the velocity vector  of the  $i^{th}$ swarmalator~\cite{csg2019}, while in the context of social interactions, the alignment of opinion dynamics could, in general, be multidimensional \cite{csg2019,kkar2021}. 

\noindent  
The first and second terms  in Eq.~(\ref{space})  correspond to the spatial attraction and repulsion. 
Spatial attraction between the swarmalators depends on the degree of orientation and  the parameter $J$. 
The repulsive interaction is essential to maintain the minimum separation between agents. The nature of the distance-dependent spatial interactions can be tuned
with the exponents $\alpha$, $\beta$, and $\gamma $.  The distant-dependent kernels in Eq.~(\ref{space}) act like  a Van der Waals interaction for $\beta>\alpha$ that
ensures the long-range attraction and short-range repulsion. $\bm{\omega}_i$ is the anti-symmetric angular velocity  $D\times D$ matrix of the $i^{th}$swarmalator, which 
can be represented in 3D as
\begin{equation}
\bm{W}_i=
\begin{pmatrix}
0 & -\omega_{i,3} & \omega_{i,2}\\ 
\omega_{i,3} & 0 & -\omega_{i,1}\\ 
-\omega_{i,2} & \omega_{i,1} & 0 
\end{pmatrix},
\end{equation}
where $\omega_{i,1},\omega_{i,2}$ and $\omega_{i,3}$ represents the components of the angular velocity.
The coupling strength $K_{ij}$ is given as
$$
    K_{ij}= 
    \begin{cases}
    \varepsilon_a/N_i   & \text{for } \left | \bm{x}_j-\bm{x}_i \right | \leq R, \\
     -\varepsilon_r/N_r & \text{for } \left | \bm{x}_j-\bm{x}_i \right | >R, 
    \end{cases}
$$
where $\varepsilon_a$ is the attractive phase coupling strength, $\varepsilon_r$ is the  repulsive phase coupling strength, 
 $R$ is the vision radius, $N_i$ is the number of swarmalators inside the vision sphere of the $i^{th}$ swarmalator excluding it
 (Refer to the  supplementary text~S1 and fig.~S1 for derivation of Eq.~(\ref{phase})). $\bm{\xi}^{\bf{x}}_i(t)$ and  $\bm{\xi}^{\bm{\sigma}}_i(t)$ are the
 Gaussian white noise with zero mean and strengths  $D_{x_k}$ and $D_{\sigma_k}$ characterized by 
 $\langle\bm{\xi}_i^{x_k}(t)\vert \bm{\xi}_i^{x_k}(t^\prime)\rangle=2D_{x_k}\delta(t-t^\prime)$ and 
  $\langle\bm{\xi}_i^{\sigma_k}(t)\vert \bm{\xi}_i^{\sigma_k}(t^\prime)\rangle=2D_{\sigma_k}\delta(t-t^\prime)$, respectively, where $k = 1,2,3, \ldots ,D$.
Note that $\bm{\sigma_i}$ has to be normalized at each time step to ensure it to be unit vector because of $\bm{\xi}_i^{\sigma_k}(t)$.
Swarmalators  in most real-world swarms only exchange interactions with its $N_i$-nearest neighbors that are within their sphere of influence
resulting in the notion of vision radius. Swarmalators within the vision sphere tend to align their internal state, whereas the others  $N_r=N-N_i-1$ have 
the natural tendency to repel each other.

We discuss the following more important situations, while others are presented in the supplementary material. Refer  to the methods section  
for details on simulation and parameters.
\subsection{Competitive interaction}
First we intend to exclusively unfold the influence of the competitive  attractive and repulsive interactions among the orientation vectors on the 
intriguing  self-organizing dynamics and therefore we fix  $\varepsilon_a=\varepsilon_r=0.5$. Some of the fascinating self-organized
  convergent  multistable symphonies by the swarmalator collectives 
are depicted and demarcated in Fig.~1.  Swarmalators  with the angle of inclination
$\rho_{ij}\in(\pi/2,\pi)$ are strongly attracted for $J<0$ and hence the collectives display a static async (SA) for small $R$, as the majority of
the swarmalators lie outside $R$ with the  tendency to repel each other. Swarmalators with $\rho_{ij}\in(0,\pi/2]$  are attracted  strongly for $J>0$
and exhibit SA for small $R$ and $J$ (fig.~S10, text~S4). Nevertheless, swarmalators with nearby $\bm{\sigma}$ are strongly attracted above appreciable $J$, even for small $R$, 
to self-organize to display a phase wave, which is active  (APW) due to the competitive repulsion among the orientation vectors
and weak spatial attraction as the majority of them lie outside $R$.

$N_i$  increases progressively proportional to $R$  resulting in the manifestation of multi-clusters (MC) from APW as $R$ is increased, which eventually merges together
to manifest as a single static synchronized  (SS) cluster above a large $R$. The sufficient condition for synchronization can be obtained as $R>1/\sqrt{1-J}$.
Refer to the supplementary text S5  for the detailed derivation.
Refer Table~S1 for the description of the acronyms for observed states.

Now, each cluster in MC becomes sparse as $J$ is decreased in the intermediate range of $R$, owing to a low degree of spatial attraction, 
and eventually the MC gather together 
with their preferred orientation to showcase  spiky states (SP).  Two such spiky states, namely twisted and flower states, are depicted
in Fig.~1 (fig. S14A), where the orientation vectors are radially pointed outwards from the axis of symmetry
 in the flower state and vice versa in the twisted state. Note that
the emergence of SP states are extended even for $J<0$, though sparse than those for $J>0$, as there lies a net positive spatial attraction 
for small $\vert J\vert$ and hence there is a meager local synchronization for the SP state  to persist. Further decrease in $J$, in the same range of $R$, 
facilitates an active core static phase wave (ACSPW) with
a turbulent core and the outer shell as the SPW.  
A strong attraction among the $N_r$ swarmalators for $J<0$ manifests the asynchronous core, while the synchronized 
swarmalators within $R$ are weakly attracted
leading to the SPW.  Now, the swarmalators in the core that fall within $R$  tend to synchronize
and eventually repell outside of $R$ to get asynchronized, which are again attracted,  both due to $J<0$, reinforcing the effect resulting in the active core.
An increase in $R$ for $J>0$ from SP increases $N_i$ resulting in  the synchronized core and the rest $N_r$ swarmalators form a SPW shielding the core,
such a coexisting of coherent and incoherent domains are known as chimera (CH). The coherent core increases with $R$ and eventually CH manifests as SS for a large $R$.  CH and SS transforms to a turning tube (TT) for  $J<0$,  as  the spatial attraction among  the incoherent domain is stronger, which remains rolling with $N_i$  like the active core in ACSPW.

Order parameters delineating the dynamical transitions as a function of $R$ for three distinct $J$ are depicted in fig.~S6. It is important to emphasize that the observed
self-organizing behaviors are robust against Gaussian noise (fig.~S12).
 Emerging dynamical behaviors for the attraction(repulsion) dominated
competitive interaction are depicted in fig.~S13(S15).  Phase diagrams with $\bm{\omega}_i =0$, two, and distributed orthogonal angular frequencies 
are respectively presented in figs.~A B, and C of figs.~S13-S15. Refer to the supplementary text S6-S8 for discussions. 
The heat maps of the employed order parameters corresponding to figs.~S13A-S15A are shown in fig.~S16. 

\subsection{Extreme $R$ and local attractive coupling}
Swarmalator collectives exclusively display SS(SA) for $N_i=N-1 \, (N_r=N-1)$ as all of them experience only attractive(repulsive) phase coupling.
Nevertheless, the collectives exhibit alluring patterns for exclusive local attractive coupling  among the orientation vectors as a function of $R$ 
especially for $J<0$ (see Fig.~2). Here,  we uncover a transition from SA to SPW   in contrast to the transition from SA to APW  in the competitive  interaction
as  a function of $J$ in the low range of $R$ as the influence of spatial proximity is absent on the swarmalators that  lie outside $R$. SPW
manifests as SS via MC as R is increased for $J>0$. There is a transition from SA to SS  via mixed synchronized state (MSS)   
as $R$ is increased for $J<0$.  SS  becomes more and more dense(sparse) for $J>0 \, (J<0)$  (Figs.~2A to 2D) as the spatial 
attractive coupling strength increasingly becomes stronger(weaker) as  $J$ is increased (decreased).
MC of similar sizes are  formed for  $R =0.5$ with a weak spatial attraction within the clusters
and a strong spatial attraction among the clusters resulting in the MSS.  The size of some of  the synchronized clusters increases with $R$ that are spatially 
sparse (see Figs. 2F-2G).  See fig.~S17(S18) and text~S9(S10) for  the comparison with two, and distributed orthogonal angular frequencies for $\varepsilon_r=0 \, (N_r=N-1)$.

\subsection{Competitive interaction with quenched disorder}
Next, we explore the effect of quenched disorder, $\bm{W}_i  \bm{\sigma}_i$, on the swarming dynamics due to the competitive interactions among the orientation vectors. 
We  consider equally distributed orthogonal angular frequencies $\bm{\omega}_1=[1, 0, 0]$ and $\bm{\omega}_2=[0, 1, 0]$  
for sustained precession of the orientation vectors.   The swarmalators quench their precession (movie~S8) leading to non-chiral collective states
as  in Figs.~1 and 2 for other choices of $\bm{\omega}_{1, 2}$.

Effectively, $\bm{W}_i  \bm{\sigma}_i$  in Eq.~(\ref{phase}) induces  a dispersion among the orientation vectors
that lead to distinct chiral states with precessing swarmalators (see Fig.~3). The influence of $R$ and $J$ are similar to those discussed in Fig.~2.
For low values of $R$,  disordered spin (DS) manifests as spinning spiky (SSP) states above a critical value of $J>0$. From SSP, 
a synchronized spinning  state (SSS) is formed  via a multi-cluster bouncing spin (MCBS) state as $R$ is increased. A pumping state (PS) mediates the 
transition from DS to SSS.  The density of SSS decreases as $J\rightarrow -1$ as in Fig.~2. 
Precessing orientation vectors recursively results in their coherence and decoherence, which dynamically establishes dense  and 
sparse synchronized clusters, respectively. The dense clusters repel each other, whereas the swarmalators in the sparse clusters that 
fall within their $R$ are synchronized resulting in the reinforcement of MCBS for $J>0$.
A similar mechanism underlies the onset of PS for $J<0$, where recursive coherence and decoherence result in sparse synchronous and dense asynchronous
collectives dynamically resulting in the PS. \\

\subsection{Real-Worlds Systems}
Now we show that our extended model~(1) is indeed able to capture the following important  real-world swarmalators.
\subsubsection*{Schooling of Fish}
We display the defensive manoeuvre of  a real
school of fish by including self-propelling velocity and modifying the repulsive interaction 
among the orientation vectors to include the centripetal inclination of the fishes towards their center of mass to
 evade predation (see supplementary text~S11 for model description).
We have used the experimental data \cite{katz2021fish} to depict the snapshots of crystal and milling behavior~\cite{fish_shape,fish2012} of a school of fish in Figs.~4A and 4C.   
Self-organizing dynamics of our model very well mimic the observed crystal and milling behaviors  as depicted in Figs.~4B and 4D, respectively. 
Some more rich behaviors of a school of fish can be found in fig.~S19. Phase diagrams and heat maps of order parameters
are depicted in figs.~S20-S21.
The synchronization $S$ and  the spatial vorticity $\Gamma_x$ order parameters, defined in the supplementary text~S11, for both  the experimental  and 
the simulation data  (see  movies~S25 (simulation)  and S26 (experiment)
depicting the evolution of  the dynamical states and the order parameters) are shown in 
Figs.~4E and ~4F, respectively.  The null value of $S(\Gamma_x$) and  unit value of $\Gamma_x(S)$ corroborates the milling(crystal)
 behavior for $t>300 (t>700)$. The striking similarities of $S$ and $\Gamma_x$ for both the experimental and  the model data establish the
 significance of our model in predicting excellently the dynamics of a school of fish both qualitatively and quantitatively. 

\subsubsection*{Traveling waves of gene expression}
Embryonic stem cells exhibit traveling phase wave, triggered by genetic oscillators, are suspected as the key to the puzzle of constant 
vertebrae segment numbers of mouse embryonic cells even when the embryonic size is reduced~\cite{LVTC2013}.  
 In vivo fluorescence image and kymograph of LuVeLu activity in mouse embryo are shown in Figs.~4G  and~4H, respectively.
 Analogous patterns exhibited by the model (1) with the  local spatial interaction  are depicted in Figs.~4I  and~4J, respectively.
 The normalized mean intensity of the fluorescence exhibiting traveling phase wave  and  
the corresponding simulation results are presented in Figs.~4K and 4L, respectively.   The supplementary  movie~S27 displays the traveling phase wave
exhibited by the swarmalator collectives, which has  striking resemblances with the in vivo real-time imaging of genetic oscillations found in the
supplementary video 1 of Ref.~\cite{LVTC2013}. 
These remarkable similarities elucidate that our generalized model  provides valuable insights on the embryonic pattern formation.

\subsubsection*{Embryonic Cell Sorting}
Embryonic cell sorting is a process, occurring during embryonic development, in which cells spontaneously sort and aggregate to form tissue patterns~\cite{Maria2008}
 based on their cell type and adhesion properties. 
 In vitro fluorescence image of initially dissociated embryonic cells in Fig.~5A~\cite{FOTY2005} sort themselves and aggregate together
 to form tissue patterns as in Fig.~5C. Analogously, random distribution of three populations of swarmalators in Fig.~5B, 
 see supplementary text~S12 for model description,
self-aggregate into organized populations (tissue layers) in accordance with their adhesive nature (see Fig.~5D).
 The supplementary  movie~S28 displays the aggregation of  swarmalators mimicking cell sorting.
 More precise modeling is possible by incorporating further details, such as cell division, cell death and cell differentiation.

\subsubsection*{Microrobot Collectives}
Several experimental studies  illustrated the self-organizing collective states of microrobots, which has potential medical and environmental applications~\cite{wwgg2022,gardi2022}.
The increasingly sparse  static sync state (see Fig.~2 for  $J<0$) is
depicted in Figs.~5F, 5J and 5N  resembles the spatial patterns of the spinning magnetic micro-disks in Figs.~5E, 5I and 5M~\cite{wwgg2022}.
We strongly believe that the rich self-organizing patterns exhibited by the minimalistic swarmalator model~(1)  can enhance the utility of microrobots.

\subsubsection*{Aggregation in Dicyostelium discoideum}
Dicyostelium discoideum is a cellular slime mold with  unusual life cycle. The separately existing single-celled amoebae
 form multicellular structures in response to the environmental stress~\cite{gregor2010onset} (see Figs.~5G, 5K, and 5O). 
The aggregation of our swarmalator model (1) with local spatial interactions captures different life stages of the slime mold (see Figs.~5H, 5L and 5P, supplementary movie~S29).

\section{Discussion}\label{sec12}
We have proposed a D-dimensional swarmalator model and unveiled a rich variety of  multistable collective behaviors, tabulated in the supplementary Table~S1,
in the phase diagrams.  Most of which are only generated in our generalized  model (1)  and only some of them were also observed in   models discussed in the literature. 
We have  defined suitable order parameters to characterize and classify the distinct self-organized collective states.  As pointed out, the SPW
and APW  qualitatively resemble with the `asters' observed in magnetic colloids and `vortex arrays' formed by  populations of spermatoza, respectively.
Notably, spiky states have striking resemblance with the `skyrmions'  observed in magnetic materials~\cite{BIO2021}, which is a potential candidate for
future data-storage solutions and other spintronics devices.  Other detected behaviors are also promising to be identified in several real-world systems.
The qualitative resemblance will set a stage for a further deep theoretical investigation of the minimalistic swarmalator model with essential extensions.

We have provided the first evidences of strong potentials of our model. In particular, we have extended the original model to successfully capture
the schooling behavior of fishes, traveling phase wave of genetic oscillator both qualitatively and quantitatively,
embryonic cell sorting, microrobots and various life stages of  slime mold. The insights on the underlying mechanisms of self-organization of cells can be useful in
synthetic engineering tissues and organs for clinical purposes~\cite{engg2008,enggSci2018,tissueEngg}.  We strongly believe that our model 
can be used to unfold the underlying mechanism behind  self-organizing properties of micro- and nano-swimmers, self-propelling agents,
microrobot collectives, etc.  In particular, the transportation properties of  microrobot collectives can be better controlled using our model
for more precise drug delivery and other biomedical applications.  Furthermore, strategic formation by drones and precise control of their collective
functions  using our model can be used for security purposes, rescue operations, explorations, etc.  Specific interest could be studying
reconfigurable microrobots for potential applications including understanding self-healing structures.

\section*{Methods}
\subsection*{Numerical Simulations and Visualization}
We have numerically solved Eq. (1) using the Runge Kutta 4th order integration scheme with a step size of 0.1.  
Initial conditions for the position vectors are  randomly drawn from a 3D cube of length 2 with each side being uniformly distributed between [-1, 1]  
and  that for the orientation vectors are randomly drawn from the uniform distributions $\theta\in[0, \pi]$ and  $\phi\in[0, 2\pi)$. 
We have fixed $\bm{v}_i =\bm{\omega}_i =D_{x_k}=D_{\sigma_k}=0$, $\alpha=1$, $\beta=3$, $\gamma=1$, $N=100$, and   
distinct self-organizing behaviors are classified in the $(J, R)$ parameter 
space in the range of $J\in[-1,1]$ and $R\in[0,2]$ throughout the manuscript unless otherwise specified. 

Each swarmalator is represented by a 
cone  (colored according to the heat map, fig.~S1) with its apex  pointing  along the orientation vector. To test the robustness of the collective states observed in our high-dimensional swarmalator model, we have used zero mean white noise with varying noise strengths ($D_{x_k}$,$D_{\sigma_k}$) of $0.00, 0.01, 0.03$. We have observed that the self-organized states are retained despite the presence of noise (fig.~S12).

\subsection*{Order Parameters}
We have used distinct order parameters to characterize and classify the distinct self-organizing collective behaviors (see Table~S1). The synchronization order parameter quantifies the degree of coherence of the orientation vectors 
of the swarmalators, which can be defined as the norm of the average  orientation of all  the swarmalators, represented as
\begin{equation}
    S=\frac{1}{N} \left |\sum_{j=1}^N \bm{\sigma}_{j} \right |,
\end{equation}
where $ \bm{\sigma}_{j} $ is the orientation vector of the $j^{th}$ swarmalator. The synchronization order parameter $S$  varies in the range $[0,1]$. 
The  asynchronized state will have $S=0$ whereas the  synchronized state
is characterized by $S=1$. The intermediate value of $S$ between $0$ and $1$ quantify the degree of coherence among the orientation vectors.

The orientation parameter quantifies the degree of alignment of the orientation vector of the $i^{th}$  swarmalator with respect to its position vector  (fig.~S2).
The orientation parameter is used to distinguish distinct spiky states such as flower state, twisted state and star state.
\begin{equation}
\Lambda = \frac{1}{N} \sum_{i=1}^{N} \frac{\bm{x}_i.\bm{\sigma}_i}{|\bm{x}_i||\bm{\sigma}_i| },
\end{equation}
where $\bm{x}_i$ is the position vector of the $i^{th}$ swarmalator. 
The orientation parameter $\Lambda$ can vary in the range $[-1,1]$. The star state (see Table S1), in which all swarmalators have orientations pointing radially outwards from the sphere would have $\Lambda=1$. An inverted star state (in which all swarmalators would point radially inwards) would take a $\Lambda$ value $-1$.  Intermediate values of 
$\Lambda$ characterizes the degree of orientation of the swarmalators  in spiky states such as flower state, and twisted state. \\

The collective states with a spherical cavity as their core such as in spiky states and active phase wave are
characterized using the  hollowness parameter defined as
\begin{equation}
    H = \frac{\min{\{ |\bm{x}_i| \}}}{\max{\{ |\bm{x}_i| \}}} \qquad\qquad i=1,2, \dots ,N.
\end{equation}
The  hollowness parameter $H$  can vary between 0 and 1. Solid spherical states such as static phase wave, synchronized states are characterized by $H=0$.
Intermediate values of $H$ between $0$ and $1$ quantify the degree of spherical cavity forming the core of the collective states. Asymptotically active states are characterized using the kinetic energy parameter  defined as
\begin{equation}
    K = \frac{1}{N} \sum_{i=1}^N \bm{\dot{x}}^2_i.
\end{equation}
The nonzero values of $K$ indicate that the collective state is dynamic, whereas near zero values indicate that the collective state is static.  The kinetic energy parameter
is depicted as a function of time for the spinning cluster, active phase wave and static synchronized state in fig.~S3.\\

We have used K-means clustering method to identify the number of clusters $(N_c)$ using the spatial and orientation data. For implemention of this algorithm, `kmeans' function from the scikit-learn python library is used. K-means algorithm segregates the swarmalators into K clusters by iteratively minimizing the clustering error$(\xi)$. 
\begin{equation}
    \xi = \sum_{K^\prime=1}^{K}\sum_{i=1}^{N_{K^\prime}}\left|\bm{X}_{i}^{K^\prime}-\bm{\mu}^{K^\prime} \right|^2
\end{equation}
where $\bm{\mu}^{K^\prime}$ is the center of $K^\prime$ cluster. The elbow point($K_{best}$) gives the estimate for number of cluster. Refer to supplementary text S2 for more details about K-means clustering approach.

\backmatter

\section*{Declarations}

\begin{itemize}
\item \textbf{Funding} : This project was supported by the DST-SERB-CRG Project under Grant No. CRG/2021/000816.
\item \textbf{Conflict of interest/Competing interests}: Authors declare that they have no competing interests.
\item \textbf{Consent for publication} : All authors declare consent for the publication of the preseted work as research article.
\item \textbf{Availability of data and materials} : All of the data are available in the main text or the supplementary materials. The custom codes used for simualtion are avaiable upon request to the corresponding author.
\item \textbf{Authors' contributions} :\\
Conceptualization: DVS, VKC, WZ\\
Methodology: DVS, AY, KJ, VKC, WZ, JK\\
Investigation: DVS, AY, KJ, VKC, WZ, JK\\
Visualization: AY, KJ\\
Supervision: DVS, JK\\
Writing - original draft: DVS, AY, KJ, VKC, WZ, JK\\
\end{itemize}

\bmhead{Supplementary information}
This article has following accompanying supplementary information. \\ 
Supplementary Text\\
Figures S1 to S21\\
Table S1\\
References \textit{(1-7)}\\
Movies S1 to S29\\

\bmhead{Acknowledgments}
The work of V.K.C. is supported by DST-CRG Project under Grant No. CRG/2020/004353 and VKC wish to thank DST, New Delhi for computational facilities under the DST-FIST programme (SR/FST/PS- 1/2020/135) to the Department of Physics. DVS  is supported by the DST-SERB-CRG Project under Grant No. CRG/2021/000816.\\

\bibliography{ms}

\clearpage
\section{Figures}\label{sec6}

\begin{figure}[b]
    \centering
    \includegraphics[width=\textwidth]{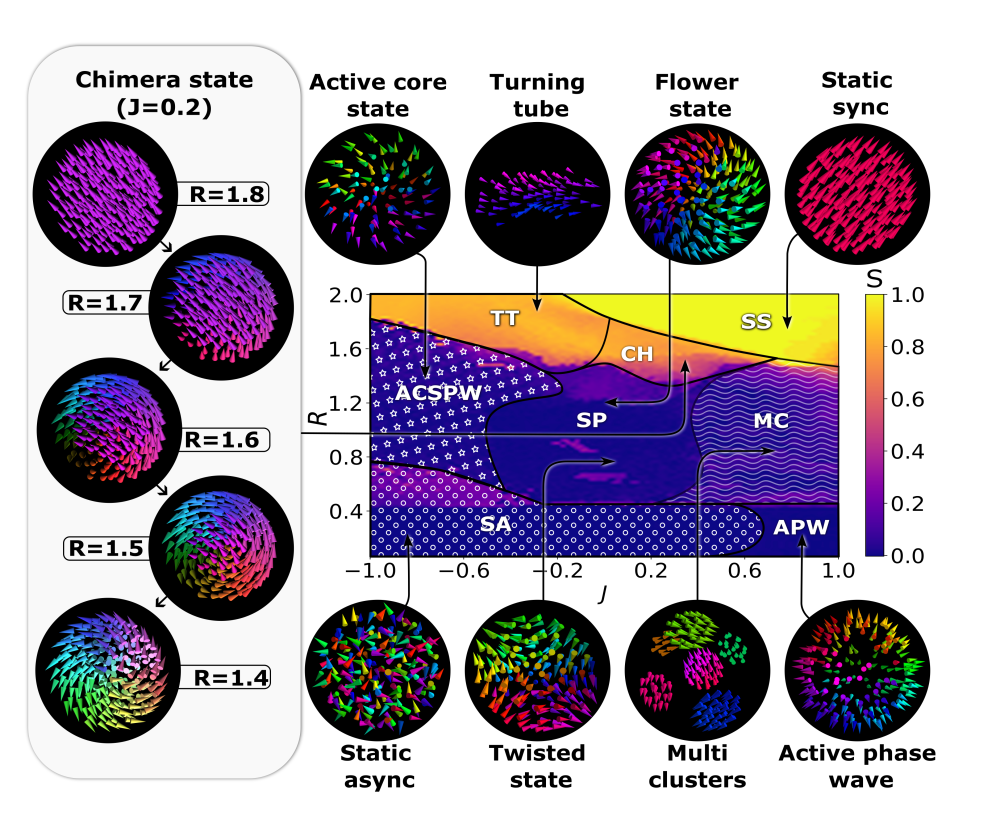}
    \caption{Synchronization order parameter $S$ is depicted as the heat map for $\varepsilon_a=\varepsilon_r=0.5$ in the $(J, R)$ space. Refer to the text for details.}
    \label{fig1}
\end{figure}

\pagebreak
\begin{figure}
    \centering
    \includegraphics[width=\textwidth]{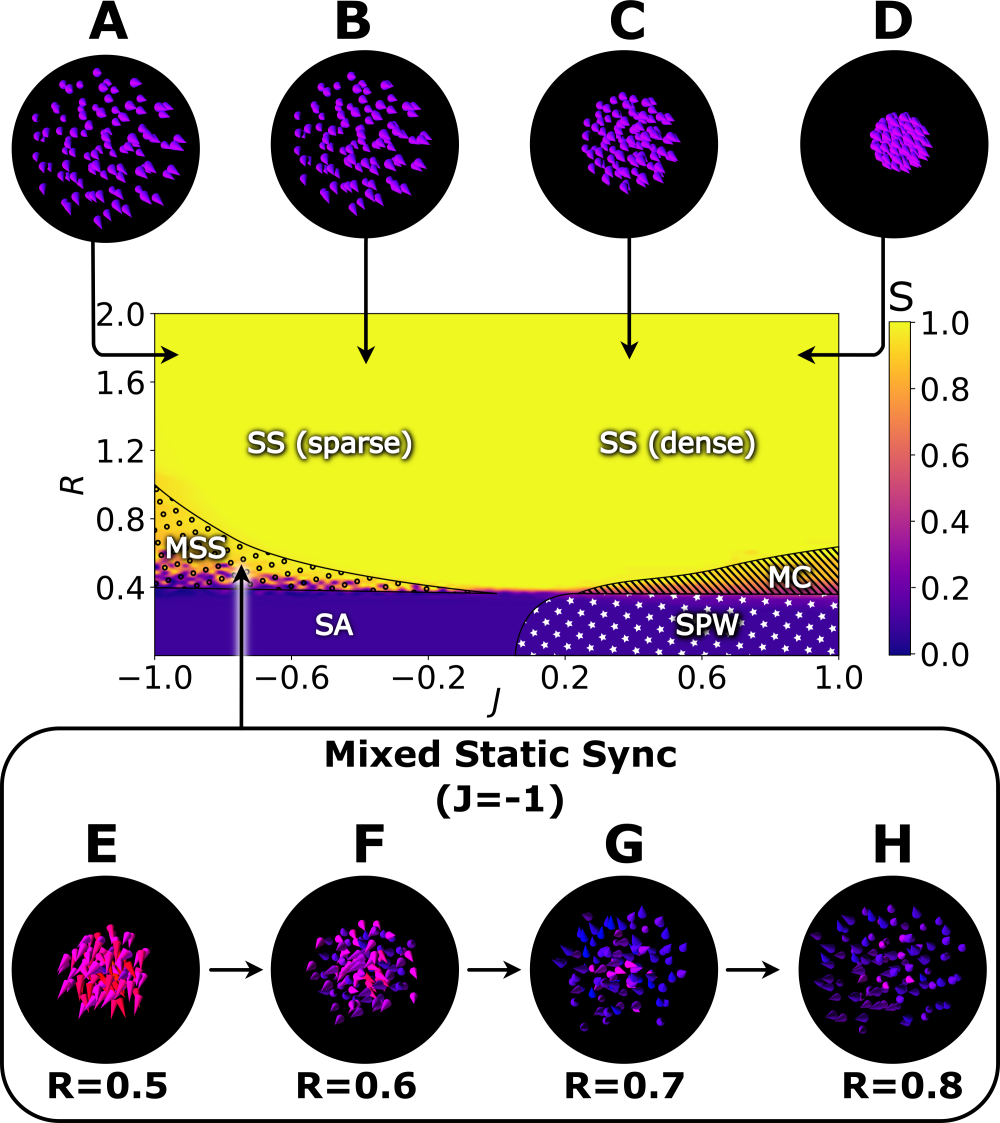}
    \caption{Synchronization order parameter $S$ is depicted as the heat map for $\varepsilon_a=0.5$, and  $\varepsilon_r=0$ in the $(J, R)$ space. Refer to the  text for details.}
    \label{fig2}
\end{figure}

\pagebreak  
\begin{figure}
    \centering
    \includegraphics[width=\textwidth]{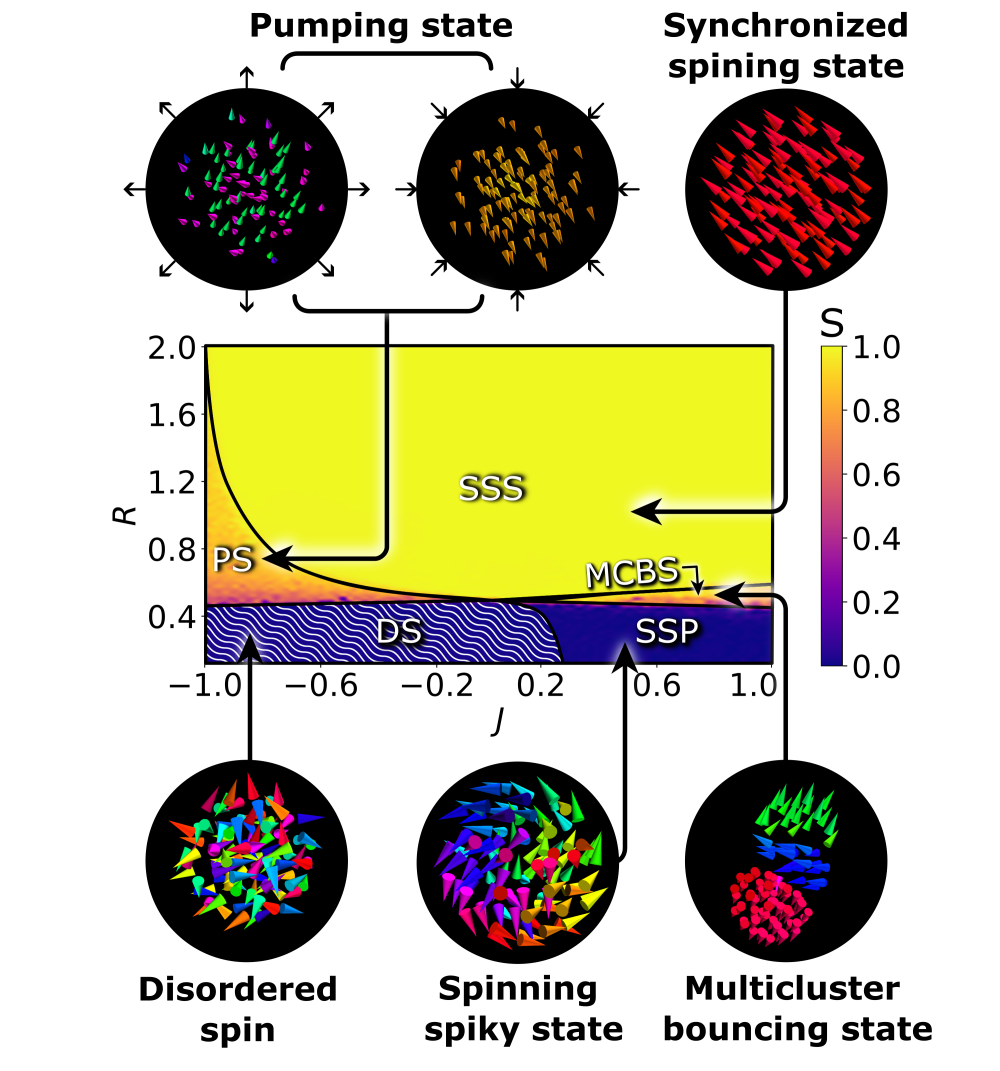}
    \caption{\textbf{Swarmalators with orthogonal angular frequencies.} Simulations are performed for $N=100,\varepsilon_a=0.9,\varepsilon_r=0.1$. The pumping state is a dynamic state in which swarmalators compress and expand in a rhythmic pattern reminiscent of the beating of the heart.}
    \label{fig3}
\end{figure}

\pagebreak
\begin{figure}
    \centering
    \includegraphics[width=\textwidth]{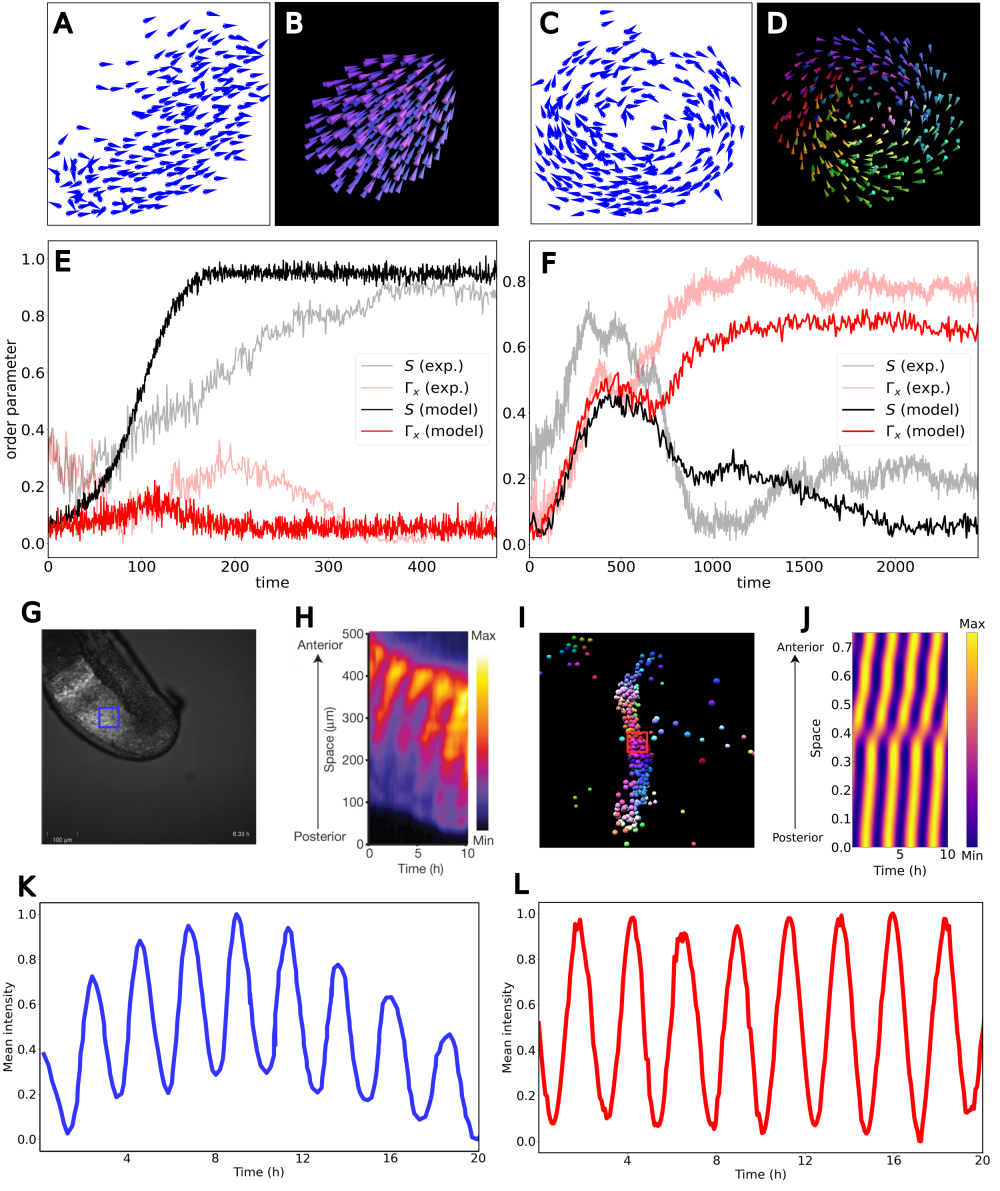}\\
    \caption{ \textbf{Real-world parallels.}     
    (\textbf{A}) Experimentally observed~\cite{katz2021fish} polarized (crystal) state for $300$ golden shiners in a shallow water at $t=330$. 
    (\textbf{B}) Polarized state in our swarmalator model for $N=300$, $R=1.0$, $J=0.7$, $\varepsilon_a=0.9$,  $\varepsilon_r=0$. 
    (\textbf{C}) Experimentally observed milling state at $t=1450$. (\textbf{D}) Milling state observed in the proposed swarmalator model 
    for $N=300$, $R=0.2$, $J=0.3$, $\varepsilon_a=0.7$, $\varepsilon_r=0$. (\textbf{E} and \textbf{F}). 
    Time evolution of synchronization $(S)$ and spatial vorticity $(\Gamma_x)$  order parameters characterizing polarized and milling states, respectively,
    from both experimental $(S(exp)$ and $\Gamma_x(exp))$, and model $(S(model)$ and $\Gamma_x(model))$ data with noise strength ($D_{x_k} =0.005, D_{\sigma_k}=0.005$).
    (\textbf{G}) In vivo fluorescence imaging and  (\textbf{H}) kymograph of LuVeLu activity in mouse embryo, which are Fig.~1 of Ref.~\cite{LVTC2013}. 
    (\textbf{I})  and   (\textbf{J}) Analogous images to (\textbf{G})  and   (\textbf{H}), respectively, from the swarmalator model.
 Normalized mean intensity of the fluorescence
   quantifying the degree of the gene expression,  estimated within the marked region in (\textbf{G}) and (\textbf{I}), in the mouse embryo  (\textbf{K}) and  the corresponding simulation results in  (\textbf{L}).The parameters are $N=500$, $R=1.56$, $\varepsilon_a=\varepsilon_r=0$, $J=1.5$, $\bm{\omega}=[1, 0, 0]$ with box size $3.2R$.}
\end{figure}

\pagebreak
\begin{figure}
    \centering
    \includegraphics[width=\textwidth]{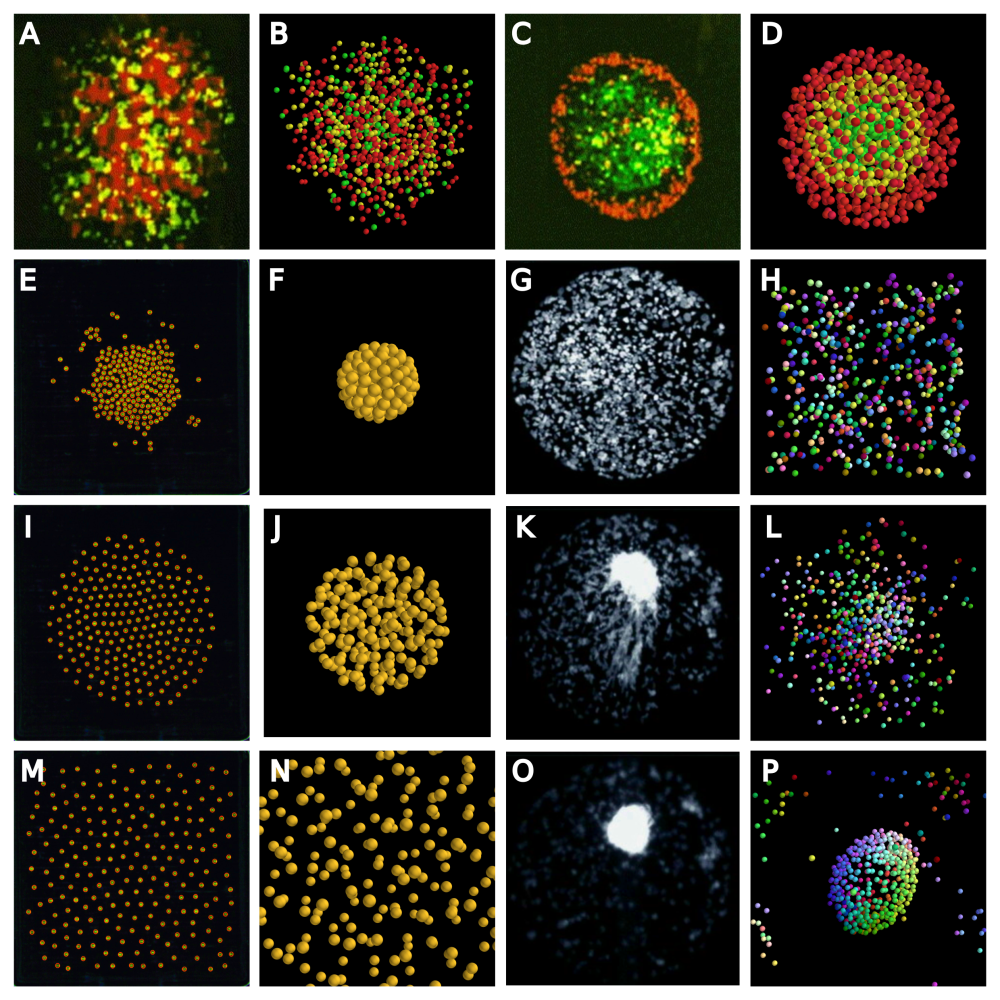}\\
    \raggedright 
    \caption{ \textbf{Real-world parallels.}   Fluorescence image of (\textbf{A})  two types of  embryonic cell mixed together and (\textbf{C}) aggregation of
    the similar types of embryonic cell, which are Figs.~4A and 4B of Ref.~\cite{FOTY2005}, respectively. (\textbf{B} and \textbf{D}) Phenomena similar to cell sorting is observed in the absence of phase couplings ($\varepsilon_a=0$, $\varepsilon_r=0$) for three populations of swarmalators. Simulation was performed with $N=900$ with different spatial coupling among swarmalator population ($J_{11}=0.9$, $J_{12}=0.7$,  $J_{22}=0.2$, $J_{23} \approx 0$, $J_{33}\approx 0$, $J_{31} \approx 0$). (\textbf{E, I, M}) Microrobot collectives expand when rotational speed is progressively increased, which are Fig.~1 of Ref.~\cite{wwgg2022}. (\textbf{F, J, N}) Increasingly sparse static sync state for  $J<0$ resembling the expanding microrobot collectives. (\textbf{G, K, O}) Snapshots illustrate the early development stages of Dictyostelium discoideum, dispersed cells self-aggregate to the final aggregation site, which are Fig.~1A of Ref.~\cite{gregor2010onset}. (\textbf{H, L, P}) The aggregation of swarmalators in case of local spatial interaction. Simulation performed for $N=500$, $R=1.4$, $J=0.8$, $\varepsilon_a=0$, $\varepsilon_r=0$ with a simulation box size of $2.1R$.}
    \label{fig5}
\end{figure}

\pagebreak
\clearpage


\setcounter{figure}{0}
\renewcommand{\figurename}{Fig.}
\renewcommand{\thefigure}{S\arabic{figure}}

\begin{center}
    {\Large Supplementary Material for \\ Exotic swarming dynamics of high-dimensional swarmalators}
\end{center}
\vspace{50 pt}
\section*{Supplementary Text}
\subsection*{S1. Derivation and features of the model}
Evolution of the orientation vectors (intrinsic dynamics) of the D-dimensional swarmalator model presented in the main text is motivated by the classical Kuramoto model. 
In higher dimensions, the orientation vector on the unit hyper-sphere is equivalent to the phase of the classical Kuramoto model. 
The orientation vectors of $i^{th}$ and $j^{th}$ swarmalators are represented as $\bm{\sigma_i}$ and $\bm{\sigma_j}$. The sine and cosine terms involving 
 angle $\rho_{ij}$ subtended by the orientation vector $\bm{\sigma_i}$ on $\bm{\sigma_j}$ can be expressed in terms of orientation vectors (Fig.~S1). In this setting,  $\cos(\rho_{ij})$  and $\sin(\rho_{ij})$ 
can be expressed as $\bm{\sigma_j}\cdot \bm{\sigma_i}$ and $\bm{\sigma_j}-(\bm{\sigma_j}\cdot \bm{\sigma_i})\bm{\sigma_i}$, respectively.\\

\noindent The notion of angular velocity has increased complexity in the higher dimensional Kuramoto model. In the classical Kuramoto model, the angular frequency represents rotation on a unit circle whereas  in higher dimensions, the orientation vector ($\bm{\sigma}$) precesses about the angular velocity vector ($\bm{\omega}$). 
The phase dynamics, that is the evolution for the orientation vector, will be governed by natural frequency in the absence of the phase coupling 
between the swarmalators ($\varepsilon_a=\varepsilon_r=0$).  \emph{Note that we use `phase dynamics'  interchangeably for `the  evolution of the orientation vector'}. 
In such a scenario,  the change in the orientation vector of a swarmalator after a time 
$dt$ can be written as $ d\bm{\sigma} = \bm{v}_{\sigma} dt $, which can be further expressed as 

\begin{equation}\tag{S1}
    \bm{v}_{\sigma}=\bm{\omega} \times \bm{\sigma}.
\end{equation}

\noindent Therefore in 3-dimensions, 
\[
\bm{\dot{\sigma}} = \bm{\omega} \times \bm{\sigma}=
\begin{vmatrix}
\hat{e}_1 & \hat{e}_2 & \hat{e}_3 \\ 
\omega_{1} & \omega_{2} & \omega_{3} \\
\sigma_{1} & \sigma_{2} & \sigma_{3} 
\end{vmatrix}
=
\begin{pmatrix}
0 & -\omega_{3} & \omega_{2}\\ 
\omega_{3} & 0 & -\omega_{1}\\ 
-\omega_{2} & \omega_{1} & 0 
\end{pmatrix} 
\begin{pmatrix}
\sigma_1 \\ 
\sigma_2 \\ 
\sigma_3
\end{pmatrix},
\]
where
\[
\bm{W}=
\begin{pmatrix}
0 & -\omega_{3} & \omega_{2}\\ 
\omega_{3} & 0 & -\omega_{1}\\ 
-\omega_{2} & \omega_{1} & 0 
\end{pmatrix}. 
\]

\noindent Now, the evolution equation for the orientation vector is given as
\begin{equation}\tag{S2}
    \dot{\bm{\sigma}} = \bm{W}\bm{\sigma}.
\end{equation}
Including the effect of the attractive and repulsive interactions of the orientation vector of the swarmalators  depending on their  vision radius $R$, 
 the evolution equation for the orientation vector corresponding to the $i^{th}$ swarmalator is governed by 
\begin{equation}\tag{S3}
\bm{\dot{\sigma}}_i = \bm{W}_i  \bm{\sigma}_i + \sum_{j=1}^N K_{ij} \left[ \frac{\bm{\sigma}_j-(\bm{\sigma}_j\cdot \bm{\sigma}_i) \bm{\sigma}_i}{\left | \bm{x}_{j}-\bm{x}_{i} \right |^\gamma} \right],
\end{equation}
where all the symbols and notations used have already been described in the model section of the main text.\\

\subsection*{S2. $K$-means clustering approach}
Clustering  is one of the interesting intrinsic feature of our model. The swarmalators self-organize to form  synchronized clusters for suitable parameters.
 We have used the heuristic Elbow curve method~\cite{elbowmethod} in $K$-means clustering  to quantify the number of clusters. \\

\noindent $K$-means clustering is an unsupervised machine-learning algorithm that segregates data into $K$ clusters~\cite{kmeans1,kmeans2}. 
The algorithm works by randomly choosing $K$ points called centroids and iteratively reassigning each data point to the cluster whose centroid is 
closer to it in terms of Euclidean distance. For further iterations, the clusters mean is taken as  their new centroid. The process is repeated until 
the Frobenius norm of the difference in the cluster centers of two consecutive iterations is less than a threshold value.\\

\noindent In particular, the swarmalators are classified on the basis of their locations and phases $X=(x_1, x_2, x_3, \sigma_1, \sigma_2, \sigma_3)$ 
into clusters $C=\{C_1, C_2, ...,C_K\}$.  
\begin{equation}\tag{S4}
    \xi = \sum_{K^\prime=1}^{K}\sum_{i=1}^{N_{K^\prime}}\left|\bm{X}_{i}^{K^\prime}-\bm{\mu}^{K^\prime} \right|^2
\end{equation}
where $\bm{\mu}^{K^\prime}$ is the center of $K^\prime$ cluster and $\xi$ is the clustering error.  We run the $K$-means clustering algorithm for our data with different values of $K$ ranging from $1$ to $20$ and evaluate the clustering error or variance. 
The intuition is that every increment in the value of $K$ will surely result in a decrease in variance $\xi$ but would have diminishing returns. At some point, when 
the value of $K$ crosses the true number of clusters, the diminish in returns will be significant enough that it can be seen as an ``Elbow'' in the $(\xi, K)$ plot (see Fig.~S4).
The elbow point,  $K_{best}$, can be numerically calculated by finding the maximum change of slope $\eta(K)$ in the $(\xi, K)$ plot. 
\begin{align*}\tag{S5}
    \eta_K&=\left[ \frac{\xi(K_{i-1})-\xi(K_{i})}{\xi(K_{i})-\xi(K_{i+1})} \right]
\end{align*}
\begin{align*}
    K_{best} &= \text{argmax}_{K}\left( \eta_K\right)
\end{align*}
The  estimation of number of clusters from the change of slope of $(\xi, K)$ plot is depicted in  Fig.~S4 for two, three and five clusters.
The distinct  clusters of  the orientation vectors maximizes their separation due to the strong repulsive interaction among the dissimilar orientation vectors
as shown in the time traces of the orientation vectors in Fig.~S5. \\

\subsection*{S3. Characterization of dynamical states using the order parameters}
Distinct self-organizing collective behaviors are depicted in  the $(J, R)$ phase diagram in Fig.~S6A, which is Fig.~1 of the main text for
 $\varepsilon_a=\varepsilon_r=0.5$, and  $N=100$. The dynamical transitions as a function of the vision radius $R$ at three different 
 values of $J=-0.9, 0.1$ and $0.9$, indicated as L1, L2 and L3, respectively in Fig.~S6A,  are depicted in Figs.~S6B to S6D, respectively.
 The synchronization order parameter $S$, hollowness $H$, kinetic energy $KE$,  number of clusters $N_c$ and orientation parameter $\Lambda$ 
 including the $K$-means clustering are  used to characterize and classify the distinct  collective dynamical states.   Note that $N_c\in(1,5)$ is 
 normalized to $N_c\in(0,1)$, so that $N_c=0$ corresponds to a single cluster,
 $N_c=0.2$ corresponds to two-cluster and so on.    In Fig.~S6B, the static asynchronous region in the range of $R\in(0, 0.375]$
 for $J=-0.9$ is characterized by the  null value of $S, KE, N_c$,  and  a very small value of $H\approx 0.1$.  
 Since the swarmalators are randomly oriented in the static async (SA) state, 
  the orientation parameter  also acquires $\Lambda=0$.  Sudden spike in the value of kinetic energy parameter at $R=0.375$  elucidates
  the active nature of the collective state `active core static spiky state' (ACSPW) in the range of $R\approx(0.375, 1.8]$. In this range of $R$,
   the values of the order parameters $S, N_c$ and $H$
  remain very low. The orientation parameter  fluctuates about  $\Lambda=0$ from positive to negative  values due to the active nature of the core 
  (see main text for explanation for the active nature of the core).  Turning tube is observed for $R>1.8$, which is characterized by a large value of 
  $S$ and $H$, whereas the parameters $\Lambda, H$ and $N_c$ acquire very low values.\\
\noindent In Fig.~S6C, there is a transition  from SA to static sync (SS) via spiky state (SP)  and chimera (CH) as a function of $R$ for $J=0.1$.  As in Fig.~S6B,  SA  in the range of $R\in(0, 0.375]$ is  characterized by near null values of all the five parameters, which manifests as SP states as $R$ is increased further. 
The spiky states consists of flower and twisted states which are hollow in nature.  For instance, see Fig.~S7D for the hollow nature of
the flower state. The hollowness parameter  acquires $H\approx 1$ in the range of $R\in(0.375, 1.4]$, while the orientation parameter 
takes some finite value.  The other parameters in this range of vision radius take very low values near zero. The SP state manifests as chimera state
for  $R\in(1.4, 1.8]$. As the latter state is characterized by coexisting coherent and incoherence domains (see Fig.~S8),  the synchronization order parameter
acquires $S\in(0.2,0.98)$ in accordance with the degree of the synchronized domain. The hollowness parameter for CH state acquires some finite but
a rather low value. As $R$ is increased further beyond $R=1.8$, CH manifests as SS state characterized by $S=1$.  The other parameters are negligibly 
small in this range of $R$.  See Fig.~S8B for the change in the
synchronization order parameter $S$ and the orientation parameter $\Lambda$ as a function of $R$ corroborating the CH and its transition to SS state.\\

\noindent In Fig.~S6D, there is a transition from active phase wave (APW) to SS via the multi-cluster (MC) state  as a function of $R$ for $J=0.9$ along L3.
KE is rather high characterizing APW.  $N_c=0.2$ elucidates that the MC is a two-cluster state, while the finite values of the other order
parameters in the MC region characterizes the nature of the cluster. For a sufficiently large $R$, MC  manifests as SS as corroborated by
a large value of  the synchronization order parameter $S$.  A schematic sketch of the distinct dynamics states observed in Fig.~S6A and the order parameters used to characterize them
for the competitive attractive and repulsive interactions between the orientation vectors is illustrated in Fig.~S9. \\

\subsection*{S4. Spatial interaction between two swarmalators}
The spatial interaction between $i^{th}$ and $j^{th}$swarmalators is governed by $F_{ij}={1+J(\bm{\sigma}_i\cdot \bm{\sigma_j})} - 1/\vert \bm{x}_{ij}\vert^2$, 
which depends on $J$ and $\bm{\sigma_i}\cdot\bm{\sigma_j}$. The $x$-intercept of the spatial interaction term $F_{ij}$ determines the equilibrium 
 distance between the two  swarmalators, which given by $1/\sqrt{1+J(\bm{\sigma}_i \cdot \bm{\sigma}_j)}$.
$J(\bm{\sigma_i}\cdot\bm{\sigma_j})\in[0,1]$ for  $J>0$ along with the angle of inclination $\rho_{ij}=\cos^{-1}(\bm{\sigma_i}\cdot\bm{\sigma_j})\in[0,\pi/2]$, and hence the
$x$-intercept of $F_{ij}$ lies between $[0.707, 1]$.  Similarly, $J(\bm{\sigma_i}\cdot\bm{\sigma_j})\in[0,1]$ for  $J<0$ and $\rho_{ij}\in[\pi/2,\pi]$. 
In both these cases, the equilibrium distance between two swarmalators lies between $[0.707, 1)$ (see Fig.~S10). \\
In contrast, $J(\bm{\sigma_i}\cdot\bm{\sigma_j})\in(-1,0]$ for  $J>0$ and $\rho_{ij}\in[\pi/2,\pi)$, and for $J<0$ and $\rho_{ij}\in(0,\pi/2]$,
in which case the  equilibrium distance between two swarmalators lies between  $[1, \infty)$ (see Fig.~S10). 
Therefore, the spatial proximity between any two swarmalators
depends on the value of  $J(\bm{\sigma_i}\cdot\bm{\sigma_j})$.  The competitive attractive and repulsive  interactions among the orientation vectors besides the vision radius  $R$
determine the value of $(\bm{\sigma_i}\cdot\bm{\sigma_j})$ for a given $J$, which in turn governs the equilibrium distance between the swarmalators.
This  underlies the reason for the manifestation of sparse states and spatially expanding states.  Further, as the two swarmalators moves out of the vision radius
to establish their equilibrium distance, their orientation vectors tend to get decoherent due to repulsive interaction between. In such a scenario, for $J<0$, the
oppositely polarized orientation vectors are attracted together to minimize their spatial separation.  The recursive  reinforcement of such an effect of  increasing  and
decreasing spatial proximity results in  breathing state, bouncing state, pumping state, and   active core state.\\

\subsection*{S5. Theoretical Analysis }
\subsubsection*{S5.1 Maximal separation between two clusters}
The maximal distance between the two-cluster state can be deduced  in the limit of large $J$ as follows.
Let $\zeta_A$ , $\zeta_B$ be the two populations of the swarmalator collectives that form two clusters $A$ and $B$, respectively.  
$N_A, N_B$ be their respective cardinal numbers.  In steady state, the mean velocity of each clusters is zero and 
consequently, the sum of the velocities of  all the swarmalators constituting  the cluster $A$ can be expressed as
\begin{equation*}\tag{S6}
\sum_{i\in\zeta_A}\bm{\dot{x}}_i =
\frac{1}{N-1}\sum_{i\in\zeta_A}\sum_{j=1}^{N} \left [ \frac{1+J (\bm{\sigma}_i \cdot \bm{\sigma}_j)}{\left | \bm{x}_{ij} \right |^\alpha} - \frac{1}{\left | \bm{x}_{ij} \right |^\beta}  \right ] \bm{{x}}_{ij}.
\end{equation*}
The second summation can be explicitly expanded as intra-cluster and inter-cluster interactions  as
\begin{equation}\tag{S7}
    \sum_{i\in\zeta_A}\sum_{j\in\zeta_A}  \left [ \frac{1+J (\bm{\sigma}_i \cdot \bm{\sigma}_j)}{\left | \bm{x}_{ij} \right |^\alpha} - \frac{1}{\left | \bm{x}_{ij} \right |^\beta}  \right ] \bm{{x}}_{ij}+\sum_{i\in\zeta_A}\sum_{j\in\zeta_B} \left [ \frac{1+J (\bm{\sigma}_i \cdot \bm{\sigma}_j)}{\left | \bm{x}_{ij} \right |^\alpha} - \frac{1}{\left | \bm{x}_{ij} \right |^\beta}  \right ] \bm{{x}}_{ij}=0.
    \label{a1a2}
\end{equation}
The first term in the above summation corresponds to the interaction within the cluster $A$, whereas the second term  in the summation corresponds to
 the interaction between the clusters $A$ and $B$.  Representing the first and second summations as $A_1$ and $A_2$, respectively,
\begin{align}\tag{S8}
A_1&=\dfrac{1}{2}\sum_{i\in \zeta_A}\sum_{j \in \zeta_A, j\neq i}\left [\bm{x}_{ij} + \bm{x}_{ji} \right ]\left [ \frac{1+J (\bm{\sigma}_i \cdot \bm{\sigma}_j)}{\left | \bm{x}_{ij} \right |^\alpha} - \frac{1}{\left | \bm{x}_{ij} \right |^\beta}  \right ]=0.
\label{a1}
\end{align}
From (\ref{a1a2}) and (\ref{a1}),  $A_2=0$.   Let $D$ be the inter-cluster distance, such that   $|x_{AB}|\approx D$.
\begin{align*}
\implies A_2=\sum_{i\in\zeta_A}\sum_{j\in\zeta_B}  \left [ \frac{1+J (\bm{\sigma}_i \cdot \bm{\sigma}_j)}{D^\alpha} - \frac{1}{D^\beta}  \right ] \bm{{x}}_{ij}.
\end{align*}

\noindent Since swarmalators within both clusters $A$ and $B$ are synchronized, $\bm{\sigma_i}=\bm{\sigma_A}$ and $\bm{\sigma_j}=\bm{\sigma_B}$ $\forall$ $i \in \zeta_A$ and $j \in \zeta_B$.

\begin{align*}
\implies A_2=\left[\dfrac{1+J (\bm{\sigma}_A \cdot \bm{\sigma}_B)}{D^\alpha}-\dfrac{1}{D^\beta}\right]\sum_{i\in\zeta_A}\sum_{j\in\zeta_B}\bm{x}_{ij}.
\end{align*}
Since $A_2=0$, and $\sum_{i\in\zeta_A}\sum_{j\in\zeta_B}\bm{{x}}_{ij}\neq0$, one can obtain the inter-cluster distance as
\begin{equation*}\tag{S9}
    D^{\beta-\alpha} = \frac{1}{ 1+ J(\bm{\sigma}_A \cdot \bm{\sigma}_B) }.
    \label{dist}
\end{equation*}
For the chosen values of the parameters  $\alpha=1$ and  $\beta=3$, the inter-cluster distance turns out to be
\begin{equation*}\tag{S10}
D=\dfrac{1}{\sqrt{1+J (\bm{\sigma}_A \cdot \bm{\sigma}_B)}}.
\end{equation*}
Maximal cluster separation is obtained when $\sigma_A\cdot\sigma_B=-1$.  Analogously,
minimal cluster separation can be obtained when $\sigma_A\cdot\sigma_B=1$.  Accordingly, the maximal and minimal separation between the two clusters are
$D_{max}={1}/{\sqrt{1-J}}$ and $D_{min}={1}/{\sqrt{1+J}}$, respectively.  

Since clusters would surely merge and synchronize once the $R>D_{max}$, the sufficient condition for emergence of static sync turns out to be
$R>{1}/{\sqrt{1-J}}$, which is numerically verified in Fig.~S11A and the analytical curve matches well with the simulation results.

\subsubsection*{S5.2 Dynamics of two swarmalators}
The equation of motion  for the spatial dynamics for the case of two swarmalators is represented as
\begin{align*}\tag{S11}
    \bm{\dot{x}}_1 &= \{ 1+ J(\bm{\sigma}_1 \cdot \bm{\sigma}_2) \} \frac{\bm{x}_2-\bm{x}_1}{|\bm{x}_2-\bm{x}_1|^{\alpha}} -\frac{\bm{x}_2-\bm{x}_1}{|\bm{x}_2-\bm{x}_1|^{\beta}}, \\
    \bm{\dot{x}}_2 &= \{ 1+ J(\bm{\sigma}_2 \cdot \bm{\sigma}_1) \} \frac{\bm{x}_1-\bm{x}_2}{|\bm{x}_1-\bm{x}_2|^{\alpha}} -\frac{\bm{x}_1-\bm{x}_2}{|\bm{x}_1-\bm{x}_2|^{\beta}}.
\end{align*}
The evolution equations governing the dynamics  of orientation vector (internal states) is given as
\begin{align*}\tag{S12}
    \bm{\dot{\sigma}}_1 &= k_{12} \left[\frac{\bm{\sigma}_2- (\bm{\sigma}_2 \cdot \bm{\sigma}_1)\bm{\sigma}_1}{|\bm{x}_2-\bm{x}_1|^{\gamma}} \right],\\
    \bm{\dot{\sigma}}_2 &= k_{21} \left[\frac{\bm{\sigma}_1- (\bm{\sigma}_1 \cdot \bm{\sigma}_2)\bm{\sigma}_2}{|\bm{x}_1-\bm{x}_2|^{\gamma}} \right].
\end{align*}

\noindent For spatially static steady states, $\bm{\dot{x}}_1=0$ and $\bm{\dot{x}}_2=0$, and  therefore the equation of motion  corresponding to  the spatial dynamics
can be expressed as
\begin{align*}
    \frac{\bm{x}_2-\bm{x}_1}{|\bm{x}_2-\bm{x}_1|^{\alpha}} \left[ \{ 1+ J(\bm{\sigma}_1 \cdot \bm{\sigma}_2) \} -   \frac{1}{|\bm{x}_2-\bm{x}_1|^{\beta-\alpha}}  \right] &=0. 
\end{align*}
Since $(\bm{x}_1 - \bm{x}_2)\neq 0$, we get
\begin{equation*}\tag{S13}
    |\bm{x}_2-\bm{x}_1|^{\beta-\alpha} = \frac{1}{ 1+ J(\bm{\sigma}_1 \cdot \bm{\sigma}_2) }.
    \label{dist_ts}
\end{equation*}
Substituting the above for the spatial separation between the two swarmalators  in  the evolution equation for the orientation vectors, the latter can be expressed  only in terms of the orientation vectors as
\begin{align*}\tag{S14}
    \bm{\dot{\sigma}}_1 &= k\left[1+ J(\bm{\sigma}_1 \cdot \bm{\sigma}_2)\right]^{\frac{\gamma}{\beta-\alpha}} \left[\bm{\sigma}_2 - (\bm{\sigma}_1\cdot \bm{\sigma}_2)\bm{\sigma}_1\right],\\
    \bm{\dot{\sigma}}_2 &= k\left[1+ J(\bm{\sigma}_1 \cdot \bm{\sigma}_2)\right]^{\frac{\gamma}{\beta-\alpha}}  \left[\bm{\sigma}_1 - (\bm{\sigma}_1\cdot \bm{\sigma}_2)\bm{\sigma}_2\right].\\
\end{align*}
The above equations completely describe the static states.  Note that Eq.~\ref{dist_ts} exactly turns out to be the separation between the two clusters $D=1/{\sqrt{1-J}}$.
Here, in the case of two swarmalators, the  sufficient condition  for synchronization is exactly the same as deduced from two-cluster state. Two swarmalators can display
static syc (SS), static asyc (SA) and static phase wave which cannot be distinguished from SA. In the region described by the 
condition ${1}/{\sqrt{1-J}}<R<{1}/{\sqrt{1+J}}$ (see Fig.~S11B), the swarmalators will have intermediate synchronization in the negative J region due to oscillations arising from the alternative synchronization and desynchronization when they enter and leave the vision radius.

\subsection*{S6. Attraction dominated competitive interaction between the orientation vectors}
Phase diagram in the $(J, R)$ parameter phase is depicted in Fig.~S13 for $N=100, \varepsilon_a=0.9$ and $\varepsilon_r=0.1$, 
to unravel the role of attraction dominated competitive interaction between the orientation vectors on the self-organizing behavior of the swarmalator collectives
for the following three distinct cases:

\subsubsection*{S6.1 In the absence of angular frequency $\bm{\omega}=0$:}
The phase diagram (see  Fig.~S13A)  for this case almost resembles the phase diagram in Fig.~2 of the manuscript for purely local attractive 
coupling without angular frequency components.   This is because of  the feeble repulsive coupling strength $\varepsilon_r=0.1$ and strong attractive coupling
strength $\varepsilon_a=0.9$, which is almost the case of local attractive coupling.
 The only difference is that the region of mixed synchronized state is replaced by the chimera  state in Fig.~S13A.  Refer
 the main manuscript for further explanation of  Fig.~S13A.
\subsubsection*{S6.2  Orthogonal angular frequencies $\bm{\omega}_1\perp \bm{\omega}_2$:}
Half of the swarmalators collectives is distributed with $\bm{\omega}_1=[1, 0, 0]$  and other half with $\bm{\omega}_2=[0, 1, 0]$.  
Note that the intra-population is homogeneous,  while the inter-population is heterogeneous with orthogonal angular frequencies.
This figure is the same as Fig.~3 of the main manuscript. It is depicted again here to appreciate the difference in the  emerging collective states
 (see Fig.~S13B), due to the angular frequency of the orientation vectors, for the case of attraction dominated competitive  interaction
 between the orientation vectors
 with that in  Fig.~S1A, where $\bm{\omega}=0$.   The presence of the angular frequency for the orientation vectors induces active states 
 in the phase diagram.  For instance, CH is replaced by  the pumping state, MC  manifested as multi-cluster bouncing state (MCBS) and
 APW appeared as spinning spiky state (SSP).  Further, the static async (SA) emerged as disordered spin (DS) state, while the  static syc (SS)
endowed with spin resulting establishing synchronized spinning state (SSS). \emph{Snapshots of all these state and others observed in the two phase diagrams 
in the following sections  are depicted in Table~1.}
\subsubsection*{S6.3  Distributed angular frequencies $\bm{\omega}_1\perp \bm{\omega}_2$:}
Half of the population has their angular frequency $\bm{\omega}_1=[\omega_1, 0, 0]$ randomly selected from the uniform distribution $\omega_1\sim U(1,3)$,  while
the other half have their angular frequency $\bm{\omega}_2=[0, \omega_2,  0]$ randomly selected from the uniform distribution $\omega_2\sim U(-1,-3)$.  Note that
the entire swarmalator collectives is characterized by heterogeneous natural frequencies.  In this case, $R\in(0, 0.4)$ has only DS in the
entire explored range of $J$ (see  Fig.~S13C).  The entire phase diagram in the region $R>0.4$ and $J>0$ is dominated by
static multi-cluster (SMC).  For $J<0$, there is a transition from the pumping state (PS) to static embedded two-cluster (static E2C) as $R$
is increased above $0.4$.  Note that
most of the spinning state observed for  $\bm{\omega}_1\perp \bm{\omega}_2$  is disappeared in this case of distributed angular frequency due to the fact that they
mutually suppress the angular precession of the orientation vectors.\\

\subsection*{S7. Competitive interaction with $\varepsilon_a=\varepsilon_r$}
Now, we will investigate the effect of competitive interaction with equal attractive and repulsive coupling strengths between the orientation vectors of 
$N=100$ swarmalator collectives with $\varepsilon_a=\varepsilon_r=0.5$.
\subsubsection*{S7.1 In the absence of angular frequency  $\bm{\omega}=0$:}
The phase diagram depicted in Fig.~S14A  is exactly Fig.~1 of the main manuscript.   It is depicted again here to appreciate the
 difference in the  emerging collective states due to  the angular frequency of the orientation vectors in the following. Kindly refer main text for explanation on
 the involved intricacies about the dynamical transitions observed in Fig.~S14A. The heat maps of order parameters used to characterize the dynamical states are shown in Fig.~S16.
\subsubsection*{S7.2  Orthogonal angular frequencies $\bm{\omega}_1\perp \bm{\omega}_2$:}
Half of the swarmalators collectives is distributed with $\bm{\omega}_1=[1, 0, 0]$  and other half with $\bm{\omega}_2=[0, 1, 0]$.  
 The swarmalator collectives result in the manifestation of several active states  upon inclusion of the orthogonal angular frequencies to the orientation vectors.
Disordered async (DA) is observed  in the range of $R\in(0, 0.4)$  for $J\in(-1.0, 0.6)$.  For $J>0.6$ active async (AA) onsets.
In the intermediate range of $R$, there is a transition from active core static spiky state (ACSSP) to bouncing multi-cluster (BMC) 
via spinning spiky state (SSP)  as  a function of $J$. For $R\approx(1.8, 2.0)$,  breathing state (BS) emerges as bouncing two-cluster (B2C)
via static embedded two-cluster (SE2C) as $J$  is increased from $-1.0$ (see  Fig.~S14B).
\subsubsection*{S7.3  Distributed angular frequencies $\bm{\omega}_1\perp \bm{\omega}_2$:}
Half of the population has their angular frequency $\bm{\omega}_1=[\omega_1, 0, 0]$ randomly selected from the uniform distribution $\omega_1\sim U(1,3)$,  while
the other half have their angular frequency $\bm{\omega}_2=[0, \omega_2,  0]$ randomly selected from the uniform distribution $\omega_2\sim U(-1,-3)$.
Here, DS prevails in the range of $R\in(0, 0.4)$ for the entire explored range of $J$. AA manifests as MC above a critical value of $J$ 
 in the range of $R\approx(0.4, 1.6)$  (see  Fig.~S14C). SE2C leads to MC for $R>1.6$ as $J$ is increased in the explored range of $J$.\\

\subsection*{S8. Repulsive dominated competitive interaction}
In this section, we will unfold the effect of repulsive dominated competitive interaction between the orientation vectors of $N=100$ 
swarmalator collectives with $\varepsilon_a=0.1$ and $\varepsilon_r=0.9$.
\subsubsection*{S8.1 In the absence of angular frequency  $\bm{\omega}=0$:}
The phase diagram depicted in  Fig.~S15A almost resembles the phase diagram in Fig.~S14A for
equal attractive and repulsive interactions between the orientation vectors without any frequency distributions. The two parameter space
corresponding to the turning tube (TT) in  Fig.~S14A display breathing state (BS) and  there is a small region of 
spinning cluster (SC)  preceding the MC  from the APW as $R$ is increased from the null value. 
\subsubsection*{S8.2  Orthogonal angular frequencies $\bm{\omega}_1\perp \bm{\omega}_2$:}
Half of the swarmalators collectives is distributed with $\bm{\omega}_1=[1, 0, 0]$  and other half with $\bm{\omega}_2=[0, 1, 0]$.  
In the low range of the vision radius $R$,   disordered spin (DS) prevails in the range of $J\in(-1,0.8)$, which manifests as AA for $J>0.8$ (see Fig.~S15B). 
In the range of $R\approx(0.4, 0.6)$ there is a transition from AMPW to BMC as a function of $J$. In the range of $R\approx(0.6, 0.1.7)$
ACSSP dominates in a rather large region of the two parameter space, which then manifests as SSP in the range of $J\approx(-0.2, 0.5)$ and
finally ending up as BMC for $J>0.5$.  There is also a transition from ACSSP  to BMC  via spinning chimera (SCH).  Finally, in the range of $R\approx(1.7, 2.0)$
there is a transition from AA$\rightarrow$SCH$\rightarrow$BMC and AA$\rightarrow$BS$\rightarrow$S2CS$\rightarrow$BMC. 
\subsubsection*{S8.3  Distributed angular frequencies $\bm{\omega}_1\perp \bm{\omega}_2$:}
Half of the population has their angular frequency $\bm{\omega}_1=[\omega_1, 0, 0]$ randomly selected from the uniform distribution $\omega_1\sim U(1,3)$,  while
the other half have their angular frequency $\bm{\omega}_2=[0, \omega_2,  0]$ randomly selected from the uniform distribution $\omega_2\sim U(-1,-3)$.
DS  prevails in the range of $R\in(0, 0.4)$ for the entire explored range of $J$.  AA(MC) emerges in almost entire range of $R$ for $J<0(>0)$,
while there lies a small region of SE2C mediating AA and MC for $R\in(1.8, 2.0)$ as a function of $J$ (see Fig.~S15C).\\

\subsection*{S9. Local attractive coupling}
The repulsive interaction between the orientation vectors are  absent when  $\varepsilon_r=0$ and hence this scenario corresponds
to purely local  coupling among the orientation vectors.  We have  fixed $\varepsilon_a=0.5$. Note that in this case, 
those swarmalators that lie within the vision radius
will have positive attraction between the orientation vectors, whereas those lie outside the vision radius lacks the influence of their spatial proximity
on the dynamics of the orientation vectors. Hence the distributed initial conditions  corresponding to the  orientation vectors that lie outside $R$
 just spatially align in accordance with the evolution equation for their position vectors.
\subsubsection*{S9.1 In the absence of angular frequency  $\bm{\omega}=0$:}
The phase diagram in the $(J, R)$ parameter space depicted in Fig.~S17A is  Fig.~2 of the main manuscript. It is depicted again here to appreciate the
dynamical change in the phase diagram when angular frequencies are included to the orientation vectors.
\subsubsection*{S9.2 Orthogonal angular frequencies $\bm{\omega}_1\perp \bm{\omega}_2$:}
Half of the swarmalators collectives is distributed with $\bm{\omega}_1=[1, 0, 0]$  and other half with $\bm{\omega}_2=[0, 1, 0]$.  
Inclusion of orthogonal angular frequencies to the orientation vectors  facilitates active states in most part of the phase diagram (see  Fig.~S17B).
SA state in Fig.~S17A  became DS.  SPW manifested as SSP, sparse SS as pumping state (PS) and very dense SS as MCBS.
\subsubsection*{S9.3  Distributed angular frequencies $\bm{\omega}_1\perp \bm{\omega}_2$:}
Half of the population has their angular frequency $\bm{\omega}_1=[\omega_1, 0, 0]$ randomly selected from the uniform distribution $\omega_1\sim U(1,3)$,  while
the other half have their angular frequency $\bm{\omega}_2=[0, \omega_2,  0]$ randomly selected from the uniform distribution $\omega_2\sim U(-1,-3)$.
In this case of purely local attraction, the phase diagram in Fig.~S17C  is exactly similar to that in  Fig.~S13C  for the
case of attraction dominated competitive interaction with distributed angular frequencies. Only difference is that the spread of the PS state in the 
phase diagram is reduced by increased in the spread of the static embedded two-cluster state.\\

\subsection*{S10. Global repulsive coupling}
All the swarmalator collectives lie outside the vision radius of $i^{th}$ swarmalator for $R=0.0$  and hence this scenario corresponds
to purely global repulsive coupling among the orientation vectors.  We have fixed $N=100$.  Now, we will illustrate the effect of the
global repulsive coupling on the swarmalator collectives in the following.
\subsubsection*{S10.1 In the absence of angular frequency  $\bm{\omega}=0$:}
Most of the phase diagram is dominated by the static async  in the $(J, \varepsilon_r)$ parameter space (see Fig.~S18A).  
However, for $J>0$ there is a transition from static async to active phase wave via static phase wave as a function of $J$ in
the range of $\varepsilon_r\in(0, 0.2)$. For $\varepsilon_r>0.2$ and  one can observe  the transition from static async to active phase wave  
above a critical value of  $J$.
\subsubsection*{S10.2  Orthogonal angular frequencies $\bm{\omega}_1\perp \bm{\omega}_2$:}
Half of the swarmalators collectives is distributed with $\bm{\omega}_1=[1, 0, 0]$  and other half with $\bm{\omega}_2=[0, 1, 0]$.  
SA in the previous case manifests as disordered spin (DS) in the presence of  two populations of swarmalators with 
mutually  perpendicular angular frequencies (see Fig.~S18B).  APW and SPW becomes spinning spiky states (SSP).
\subsubsection*{S10.3  Distributed angular frequencies $\bm{\omega}_1\perp \bm{\omega}_2$:}
Half of the population has their angular frequency $\bm{\omega}_1=[\omega_1, 0, 0]$ randomly selected from the uniform distribution $\omega_1\sim U(1,3)$,  while
the other half have their angular frequency $\bm{\omega}_2=[0, \omega_2,  0]$ randomly selected from the uniform distribution $\omega_2\sim U(-1,-3)$.
The entire phase diagram is occupied only by disordered spin because of purely global repulsion and heterogeneous nature of the 
natural frequencies of the orientation vectors (see Fig.~S18C).

*{S11. Extended model for school of fish} 
In this section, we extend our model to capture the self-organizing behaviors of  school  of fish by including the self-propulsion velocity and slightly modifying
the phase repulsive interaction.  The orientation vector describing the  internal state of a  swarmalator  can be interpreted as the heading direction of the swarmalator (fish), which is a typical characteristic feature of school of fish during their bait-ball or milling behavior~\cite{TKKY2013}. These collective behaviors offer several advantages such as enhanced
 predator defense and increased foraging efficiency ~\cite{ACCJ2013}. We choose the self-propulsion velocity along the orientation of the agent as
\begin{equation*}\tag{S15}
    \bm{v}_i=c_i\bm{\sigma}_i,
\end{equation*}
where, $c_i$ is the mapping coefficient for self-propulsion velocity. Hence the spatial dynamics in Eq.~(1) of the main text can be modified as

\begin{equation}\tag{S16}
    \bm{\dot{x}}_i = c_i\bm{\sigma}_i + \frac{1}{N-1}\sum_{j=1} ^N \left [ \{1+J (\bm{\sigma}_i \cdot \bm{\sigma}_j)\} \frac{\bm{x}_{j}-\bm{x}_{i}}{\left | \bm{x}_{j} -\bm{x}_{i} \right |^\alpha} - \frac{\bm{x}_{j}-\bm{x}_{i}}{\left | \bm{x}_{j}-\bm{x}_{i} \right |^\beta}  \right ]. 
    \label{fish_space}
\end{equation}

\noindent Bait-ball or milling of school of fish maintains their social boundary with some radius $L$ with respect to their center of mass. During schooling, fishes maneuver 
their orientation such that they stay inside this social structure $(L)$ to increase their chance of survival~\cite{CIVI2012}.  The evolution equation corresponding to the orientation vector 
can be represented as
\begin{equation}\tag{S17}
    \bm{\dot{\sigma}}_i = \bm{W}_i  \bm{\sigma}_i + \sum_{j=1}^N \frac{\varepsilon_a}{N_i} \left[ \frac{\bm{\sigma}_j-(\bm{\sigma}_j\cdot \bm{\sigma}_i) \bm{\sigma}_i}{\left | \bm{x}_{j}-\bm{x}_{i} \right |^\gamma}\right] - \bm{F}_i.
    \label{fish_phase}
\end{equation}

\noindent  Note that the $\bm{\omega}_i =0$ and the repulsive  coupling among the orientation vectors in Eq.~(1) of the main text is replaced by the term  $\bm{F_i}$, which is expressed as
\begin{equation}\tag{S18}
    \bm{F}_i = \frac{\bm{x}_i^c-(\bm{x}_i^c\cdot \bm{\sigma}_i) \bm{\sigma}_i}{\left | L\bm{\hat{x}}_{i}^c-\bm{x}_{i}^{c} \right |},
\end{equation}
where $\bm{x}_{i}^c = \bm{x}_i -\bm{x}_c$ represents the position of $i^{th}$ swarmalator with respect to their center of mass $\bm{x}_c =\sum_{i=1}^N \bm{x}_i/N$.
The term $\bm{F}_i$ can be considered as the force that  enforces the  centripetal inclination of the school of fish which facilitates them to maintain the 
dense bait-ball formation to evade their predators~\cite{WDHN1971}.\\

\noindent To describe the observed collective behaviors of school of fish, we use three order parameters, namely synchronization parameter ($S$), 
spatial vorticity ($\Gamma_x$), 
and phase vorticity ($\Gamma_\sigma$). The synchronization parameter provides the measurement of the alignment of individuals in the group.
To measure the rotational motion of the fishes, we use two vorticity parameters~\cite{TKKY2013}. Spatial vorticity ($\Gamma_x$) 
is calculated using spatial velocity and is useful for capturing the vorticity arising due to the lateral motion, while the phase vorticity is calculated using the orientation vectors
and is useful for measuring the vorticity due to motion along the heading direction. The spatial vorticity is defined as
\begin{equation}\tag{S19}
    \Gamma_x=\frac{1}{N} \left|\sum_{i}^N \bm{\hat{x}}_i^c \times \bm{\hat{v}}_i \right|,
\end{equation}
where, $\bm{\hat{x}}_i^c$  is the unit vector corresponding to the position vector of the $i^{th}$ swarmalator with respect to their center of mass $\bm{x}_c$ and 
$\bm{\hat{v}}_i$ is the unit vector of the spatial velocity of the $i^{th}$ swarmalator.
Phase vorticity is defined as 
\begin{equation}\tag{S20}
    \Gamma_{\sigma}=\frac{1}{N} \left| \sum_{i}^N \bm{\hat{x}}_i^c \times \bm{\sigma}_i \right|.
\end{equation}
Both vorticity parameters, $\Gamma_x$ and $\Gamma_\sigma$, can vary in the range $0$ to $1$.  
Typical configurations observed during the schooling of fish are the swarm state, polarized state, and milling state. In the case of milling, 
swarmalators, here fishes, show coordinated rotational motion and   in polarized (crystal) state fishes are aligned. Quantitatively, the milling state  is characterized by
 high vorticity ($\Gamma_x \approx  1, S \approx  0$), while the polarized state is characterized by the high value of
synchronized parameters ($S \approx 1, \Gamma_x \approx 0$). In the swarm state, the fish are neither ordered  nor form vorticity and hence this state is characterized by a feeble
spatial and phase vorticities ($\Gamma_x \approx 0,\Gamma_\sigma \approx  0$) and negligible value of the synchronization  order parameter ($S \approx  0$). 
The collective states observed in the modified model (Eq.~\ref{fish_space}, Eq.~\ref{fish_phase}) and the time evolution of the above order parameters are depicted
 in Fig.~S19. Heat maps of the order  parameters in $(J, R)$ space are shown in Fig.~S20 and Fig.~S21.

\subsection*{S12. Extended model for cell sorting} 
In the swarmalator model, the orientation vectors can be used to represent different types of cells.
 Similar types of cells tend to cluster together due to the different adhesive properties.  This behavior can be captured by considering two or more
  populations of swarmalators with 
 different spatial coupling. We assume that the state of the cell is not changing during the sorting process and hence $\varepsilon_a =\varepsilon_r=0$. Since $\bm{\omega}=0$,
the evolution equations of motion for three populations of swarmalators  governing the cell sorting dynamics can be given as
 \begin{equation}\tag{S21}
    \bm{\dot{x}}^{(b)}_i = \bm{v}^{(b)}_i +\sum_{b^\prime=1}^3 \frac{1}{(N^{(b')}-\delta_{bb'})}\sum_{j=1}^N \left[ 1+J^{(bb^\prime)} (\bm{\sigma}^{(b)}_i \cdot \bm{\sigma}^{(b')}_j) \frac{\bm{x}^{(b')}_{j}-\bm{x}^{(b)}_{i}}{\left | \bm{x}^{(b')}_{j} -\bm{x}^{(b)}_{i} \right |^\alpha} - \frac{\bm{x}^{(b')}_{j}-\bm{x}^{(b)}_{i}}{\left | \bm{x}^{(b')}_{j}-\bm{x}^{(b)}_{i} \right |^\beta}  \right]
\end{equation}

\noindent where $b=1, 2$ and $3$ corresponds to the  three distinct population of cells $A, B$ and $C$, respectively.   
We have assumed type-$A$ ($B$) cells have more adhesiveness than  type-$B$ ($C$) and hence accordingly,
$J^{AA}>J^{AB}>J^{BB}>J^{BC}>J^{CC}>J^{AC}=0$, where $J^{AB}=J^{BA}$, $J^{AC}=J^{CA}$, and $J^{BC}=J^{CB}$.

\pagebreak
\begin{figure*}[h]
\centering
\includegraphics[width=1\textwidth]{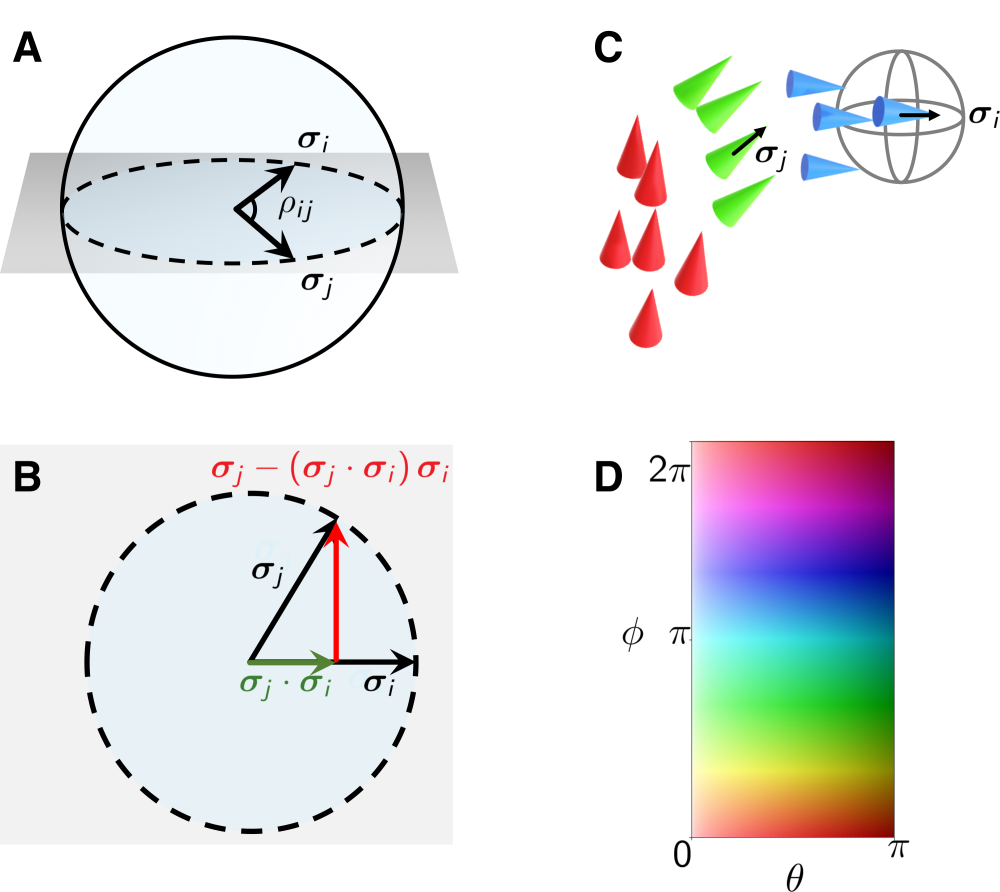}
\caption{\textbf{Deriving 3D phase mathematical model and its visualisation.} (\textbf{A}) Orientations $\bm{\sigma}_i$ and $\bm{\sigma}_j$ of $i^{th}$  and $j^{th}$ 
 swarmalators, respectively,   can be represented as  vectors pointing on a unit sphere. (\textbf{B}) Deriving the analogues of $\sin(\rho_{ij})$ and $\cos(\rho_{ij})$ in terms of orientation vectors $\bm{\sigma}_i$ and $\bm{\sigma}_j$. Here $\rho_{ij}$ is the angle of inclination of orientation vectors, $\cos(\rho_{ij})$ is simply the
 projection of $\bm{\sigma}_j$ on $\bm{\sigma}_i$, $\sin(\rho_{ij})$ is the $y$-component of $\bm{\sigma}_j$.  
  (\textbf{C}) Each swarmalator is represented by a cone  with its apex  pointing along its orientation vector.
   (\textbf{D}) The heat map, encoding the degree of orientation of the vectors which in turn is  determined by the polar angle$ (\theta)$ and 
   the azimuthal angle $(\phi)$, is used to color the cones in accordance with the distribution of 
   the initial conditions, which facilitates to identify distinct collective states even when the cones are masked behind a dense set of cones.} 
\end{figure*}

\pagebreak
\begin{figure*}[h]
    \centering
    \includegraphics[width=0.9\textwidth]{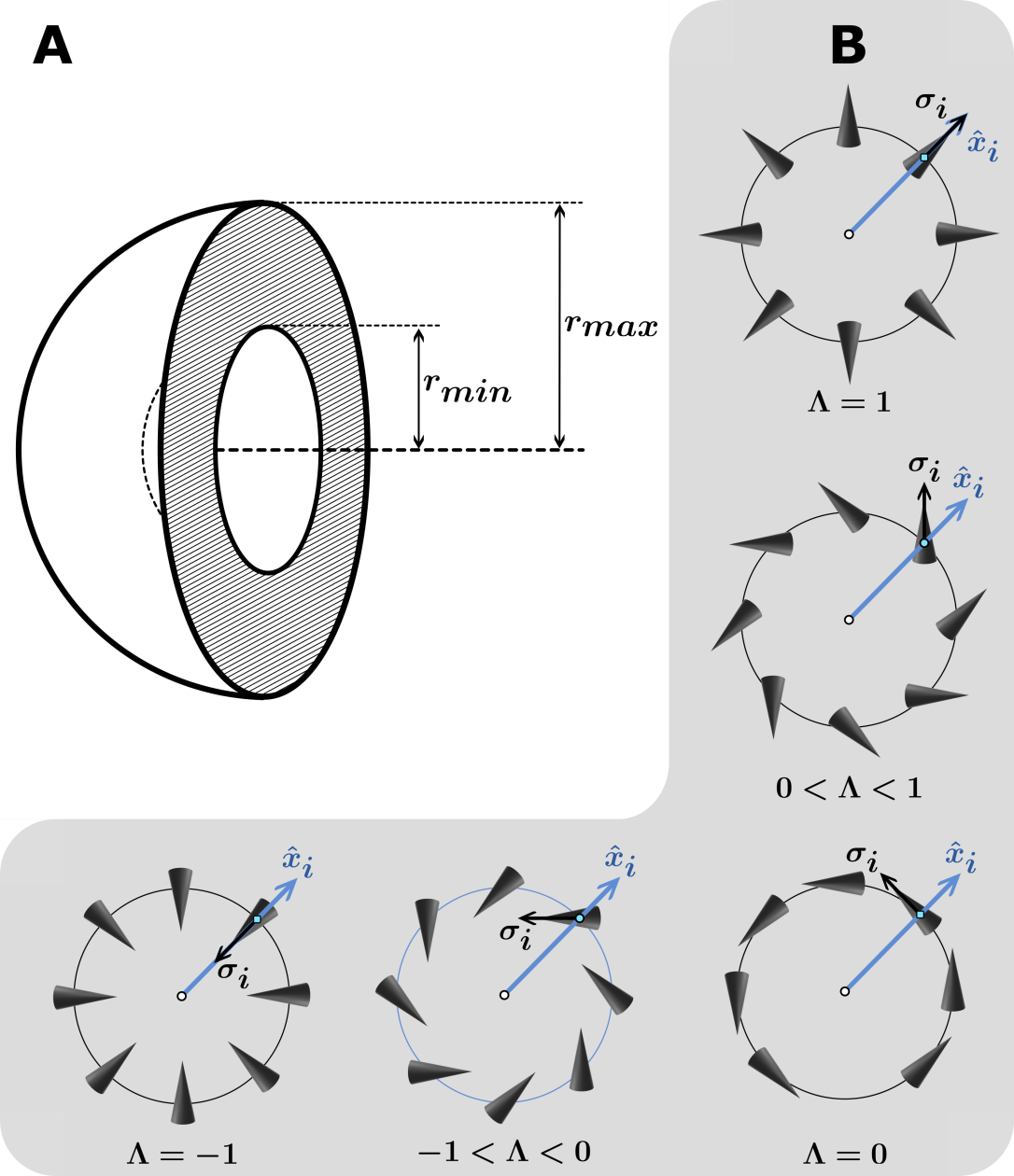}
    \caption{\textbf{Schematic diagram to facilitate the deduction of hollowness and orientation parameters.}
    \textbf{(A)} For hollow states, the radius of the hollow cavity is denoted by $r_{min}$, and the total radius of the spherical state is denoted by $r_{max}$. The  ratio between $r_{min}$ and $r_{max}$ determines the degree of hollowness. \textbf{(B)} The blue arrow $\bm{x}_i$ represents the position from the center and the black arrows $\bm{\sigma}_i$ represent the orientation of $i^{th}$ swarmalator. 
The angle between the position vector and the orientation vectors of the $i^{th}$ swarmalator determines its degree of alignment.
 }
 \label{ohp}
\end{figure*}

\pagebreak
\begin{figure}[h]
    \centering
    \includegraphics[width=1\textwidth]{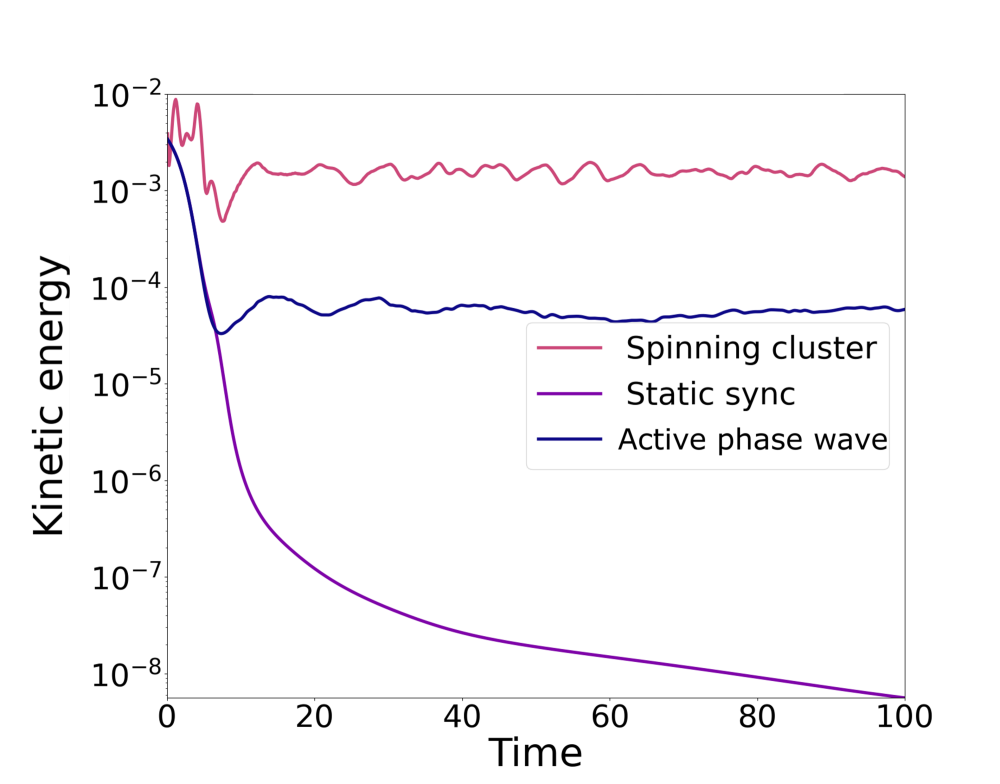}
    \caption{\textbf{Time evolution of the kinetic energy parameter.} Time evolution of the kinetic energy parameter corresponding to the static sync (SS) for $J=0.1,\varepsilon_a=\varepsilon_r=1.0, N=200$ and $r=2$, active phase wave (APW) for  $J=0.75, \varepsilon_a=\varepsilon_r=1.0, N=200$ and $R=0$, and the spinning cluster state for
    $J=1.0, \varepsilon_a=0.0, \varepsilon_r=1.0, N=200$ and $R=0.5$.  The spinning cluster can be seen in the Movie S21, as its spinning nature is difficult to perceive in the snapshot.  The initial peak in the kinetic energy parameter is due to the transient dynamics. 
     The value of the  kinetic energy parameter for the spinning cluster state is an order of magnitude greater than for the active phase wave.} 
     \label{teke}
\end{figure}

\pagebreak
\begin{figure}[h]
    \centering
    \includegraphics[width=1\textwidth]{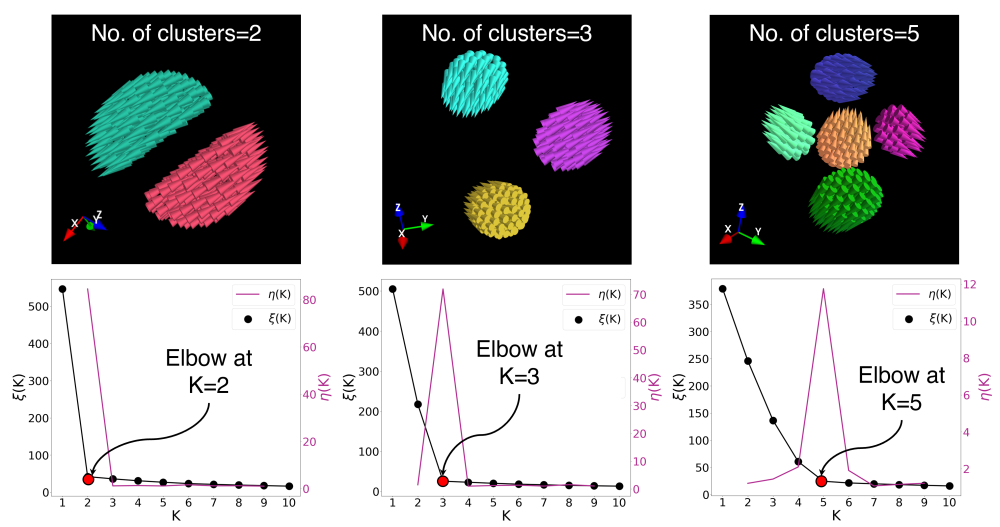}
    
    \caption{\textbf{Estimation of number of clusters using $K$-means clustering.}  The diagram illustrates the implementation of the $K$-means clustering framework
     (discussed in the supplementary text~S2) to estimate the number of clusters in the `multi-cluster states'. The  number of clusters can be found using the elbow method
      from the $(\xi, K)$ phase plot. 
     The black line corresponds to the error $\xi(K)$ whereas the red line corresponds to the change of slope function $\eta(K)$. The elbow is the point where the 
     change of slope is maximum, seen as a peak in the red curve. For $K=2$ case, half of the peak is not visible since the change of peak function is defined only for $K\geq2$. The elbow of $\xi(K)$ and the peak of $\eta(K)$ in the bottom row clearly quantify two, three and five clusters observed in the  respective column of the first row (refer  supplementary text S2 for more details).} 
     \label{kmc}
\end{figure}

\pagebreak
\begin{figure}[h]
    \centering
    \includegraphics[width=1\textwidth]{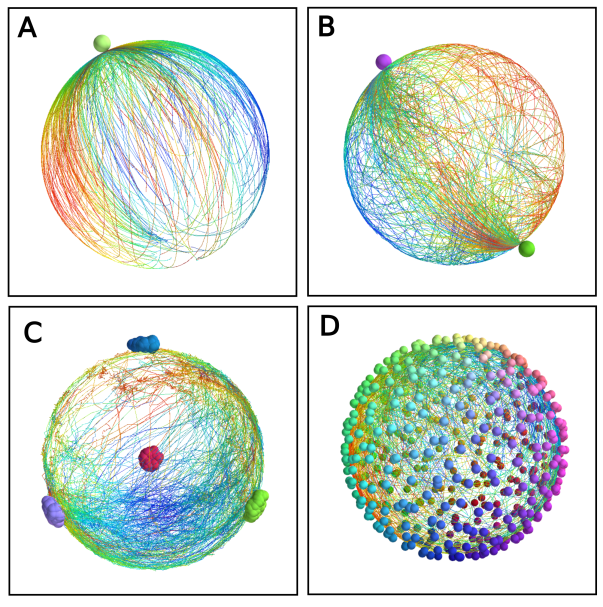}
    \caption{\textbf{Time traces of the orientation vectors.} Time traces  of  the orientation vectors corresponding to $N=200$ swarmalators. 
    The colored spheres, colored according to the heat map in Fig.~S1, are the asymptotic clusters of the orientation vectors.
    (\textbf{A}) Single cluster for  $R=2$. (\textbf{B}) Two-cluster for  $R=1$. (\textbf{C}) Four-cluster for $R=0.25$. 
 (\textbf{D}) Orientations for Twisted (Spiky) state. The values of the other parameters are $\varepsilon_a=\varepsilon_r=0.6$ and $J=0.8$. The distinct clusters of the orientation vectors maximizes their  spatial separation  due to the strong repulsive interaction among the dissimilar orientation vectors/clusters.}
 \label{ttov}
\end{figure}
\pagebreak
\begin{figure}[h]
    \centering
    \includegraphics[width=1\textwidth]{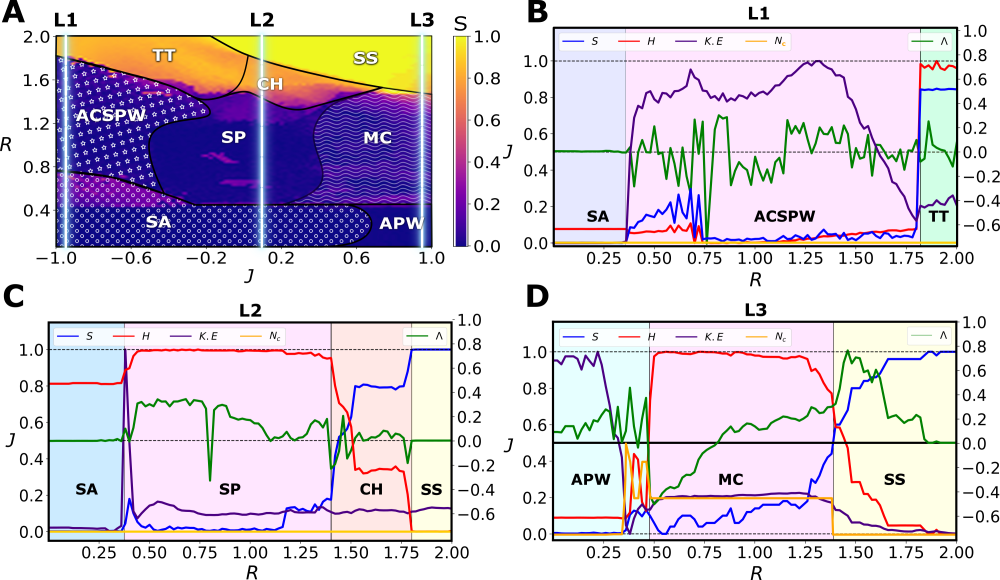}\\
    \caption{\textbf{Dynamical transition using the order parameters.} (\textbf{A}) Fig.~1 of the main text for $\varepsilon_a=\varepsilon_r=0.5$, and  $N=100$. 
    The dynamical transitions leading to distinct self-organizing collective behaviors as a function of the  vision radius $R$ at three different 
    values of $J=-0.9, 0.1$ and $0.9$ (indicated as L1, L2 and L3, respectively) are depicted in Figs.~ (\textbf{B}-\textbf{D}), respectively.
 The synchronization order parameter $S$, hollowness $H$, kinetic energy $KE$,  number of cluster $N_c$ and orientation parameter $\Lambda$ are 
 used to characterize and distinguish distinct dynamical states.  Refer supplementary text~S3 for brief explanation on characterizing the
 distinct dynamical transitions (states) using the order parameters.}
 \label{dtop}
\end{figure}

\pagebreak
\begin{figure*}[h]
\centering
    \includegraphics[width=\textwidth]{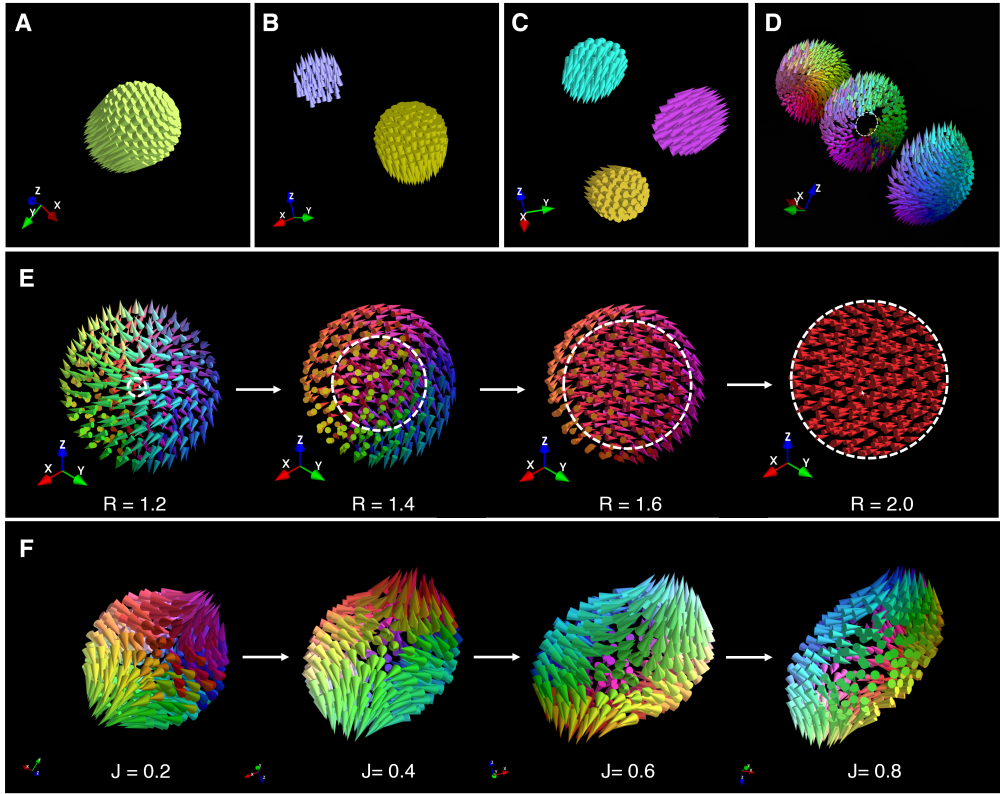}
    \caption{\textbf{Cluster states and the manifestation of chimera.}  We fix $N=500$. (\textbf{A}) Static synchronized state (single cluster)
    for $R=2$. (\textbf{B}) Two-cluster for $R=1$.  (\textbf{C}) Three-cluster for $R=0.5$. The other parameters are  $\varepsilon_a=\varepsilon_r=0.5$. (\textbf{D}) A 3D-slice plot of  Flower state to illustrate the spherical cavity at its core.  These states are  already  depicted in Fig.~1 of the main manuscript $\varepsilon_a=\varepsilon_r =0.5$. 
    (\textbf{E}) Simulation results for $J=0, \varepsilon_a=\varepsilon_r=1$. Since $J=0$  the spatial dynamics are independent of the orientation vectors of the swarmalators, but competitive  interactions among the orientation vector exists due to non-zero values of $\varepsilon_a$ and $\varepsilon_r$. 
    Manifestation of chimera  state can be  observed during the transition from the twisted state (SP) to the static sync (SS) state as the vision radius is increased. It is evident
    that as $R$ increases the coherent core increases in size enveloped by static phase wave finally resulting in completely synchronized state.
    (\textbf{F}) Flower state for $R=1.0, \varepsilon_a=\varepsilon_r=0$, the cluster elongates when the value of parameter $J$ is increased. 
    Similarly aligned swarmalators will attracted strongly in proportion to the value of $J$.
    When J is small or zero, the effect of similarity  of the orientation vectors on  the spatial proximity is absent, and the swarmalators settle on to a unit sphere. Increasing $J$ will make the swarmalators near the two poles (places where the symmetry axis passes through the sphere) come closer, hence distorting the spherical shape. 
    Some of these states are already  depicted in Fig.~1 of the main manuscript for
    $\varepsilon_a=\varepsilon_r =0.5$. Here the collective states are depicted for  $N=500$ swarmalators  and scaled up for better perception of the
    distinct self-organizing collective states.}
    \label{csmc}
\end{figure*}


\pagebreak
\begin{figure}[h]
    \centering
    \includegraphics[width=1\textwidth]{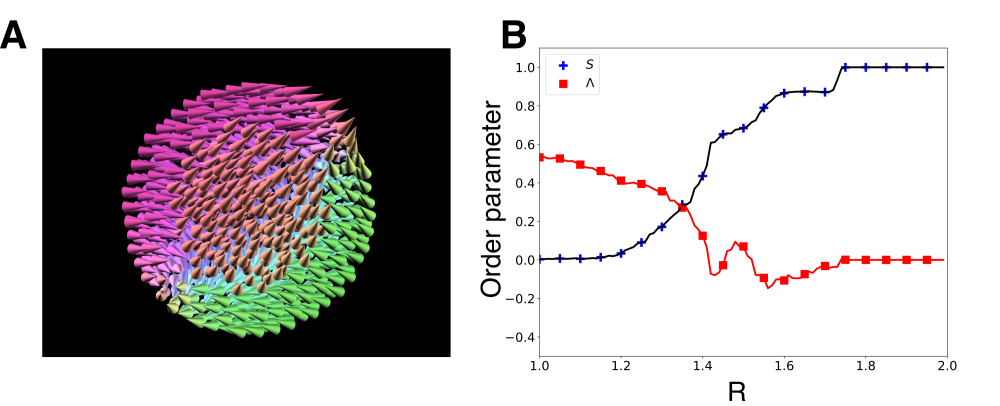}
    \caption{\textbf{The 3D cross-section of the chimera.}  (\textbf{A}) Illustration of coherent and incoherent domains constituting chimera.
    The static sync core and the phase wave shell surrounding it are clearly  illustrated. The transition between the two layers is sharp, 
    suggesting that there is an explosive transition to synchronization governed by the distance-dependent interaction.
       (\textbf{B}) The synchronization order  parameter $S$  and orientation parameter  $\Lambda$  are depicted as a function of the  vision radius $R$.
 As $R$ is increased $S$ increases characterizing the degree of synchronized core and eventually reaching unity   elucidating the emergence of SS.
 Note that  $\Lambda$ decreases as the orientation vectors align among themselves and eventually attains the null value when the static sync onsets charactering the emergence of complete coherence among the orientation vectors.}
 \label{3dcsc}
\end{figure}
\clearpage
\pagebreak

\begin{figure}[h]
    \centering
    \includegraphics[width=\linewidth]{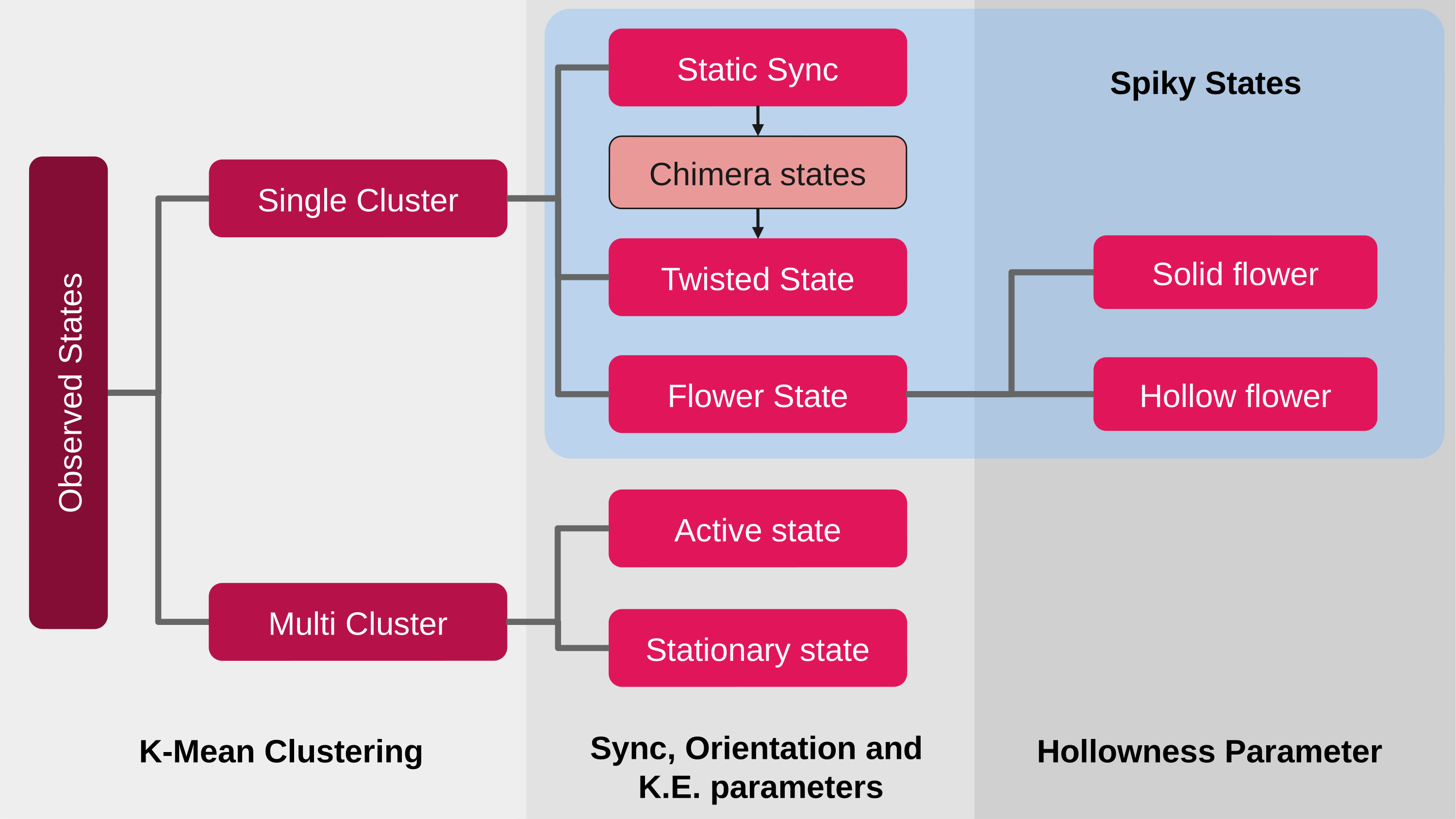}
   \caption{\textbf{Schematic sketch of the self-organizing states for competitive interaction.} 
    We have observed a variety of self-organizing collective dynamics states in our model for a different set of parameters.
     The schematic diagram shows the distinct states observed across the different ranges of parameters for  $\bm{\omega}=0$ and $\varepsilon_a=\varepsilon_r =0.5$,
     which are characterized and classified based on the values of the five distinct order parameters namely, synchronization order parameter $S$, hollowness $H$, kinetic energy $KE$,  number of cluster $N_c$ and orientation parameter $\Lambda$.}
     \label{ssdds}
\end{figure}

\pagebreak
\begin{figure}[h]
    \centering
    \includegraphics[width=1\textwidth]{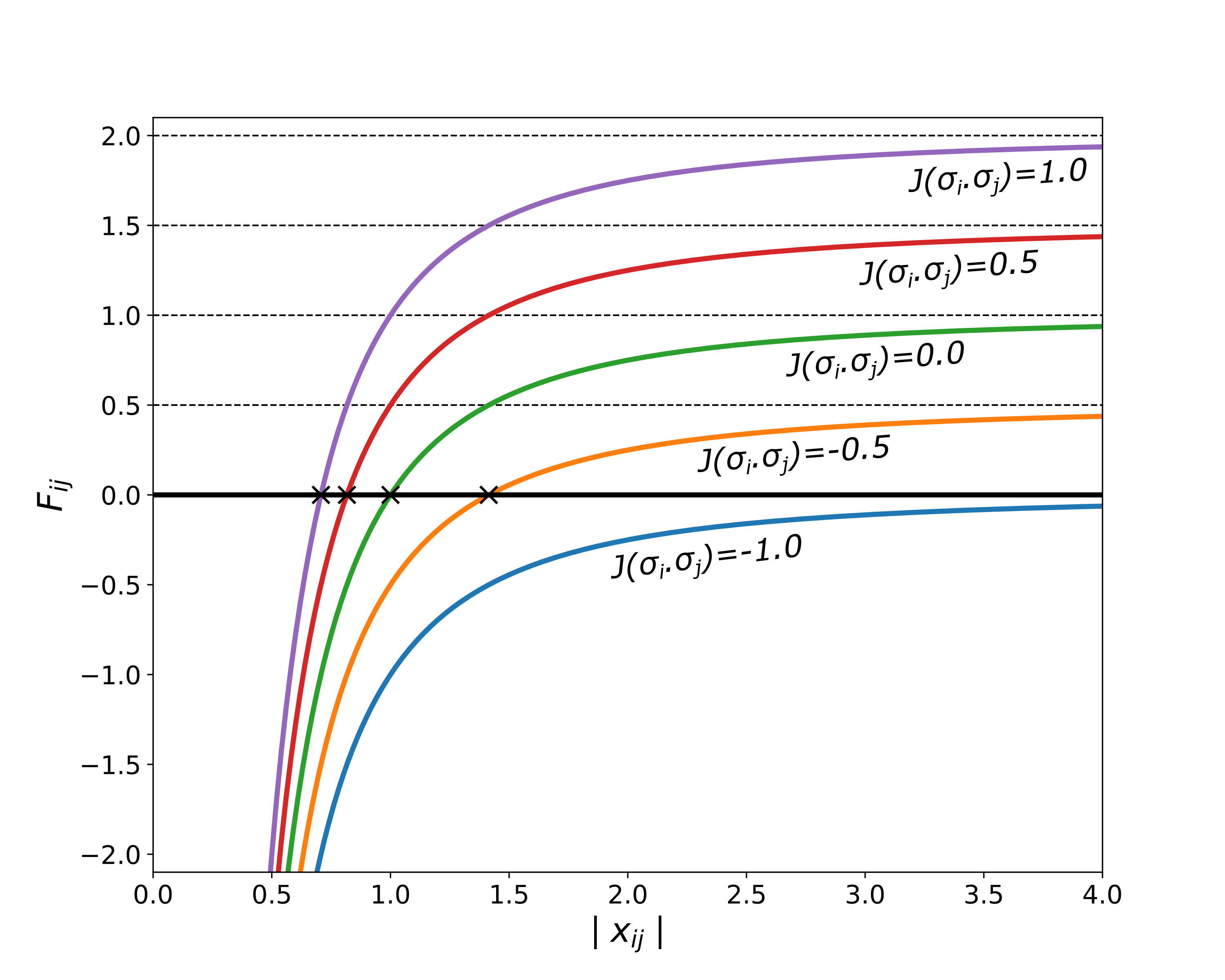}
    
    \caption{\textbf{Spatial interaction between two swarmalators $i$ and $j$.}
    The $x$-intercept of the spatial interaction term $F_{ij}=1+J(\bm{\sigma}_i\cdot\bm{\sigma}_j)-1/{\vert \bm{x}_{ij}\vert^2}$ determines the equilibrium 
      distance between the $i^{th}$ and $j^{th}$ swarmalators, given by $1/{\sqrt{1+J(\sigma_i\cdot\sigma_j)}}$. It is evident that the equilibrium distance  
      between any two swarmalators 
      can vary from $0.707$ to $\infty$ depending on the value of $J$ and the dot product of $\bm{\sigma_i}$ and $\bm{\sigma_j}$. Refer supplementary text~S4 for more
      discussions on the effect of the spatial interaction term on the observed collective dynamical states.}
      \label{sit}
\end{figure}

\pagebreak
\begin{figure}[h]
    \centering
    \includegraphics[width=1\textwidth]{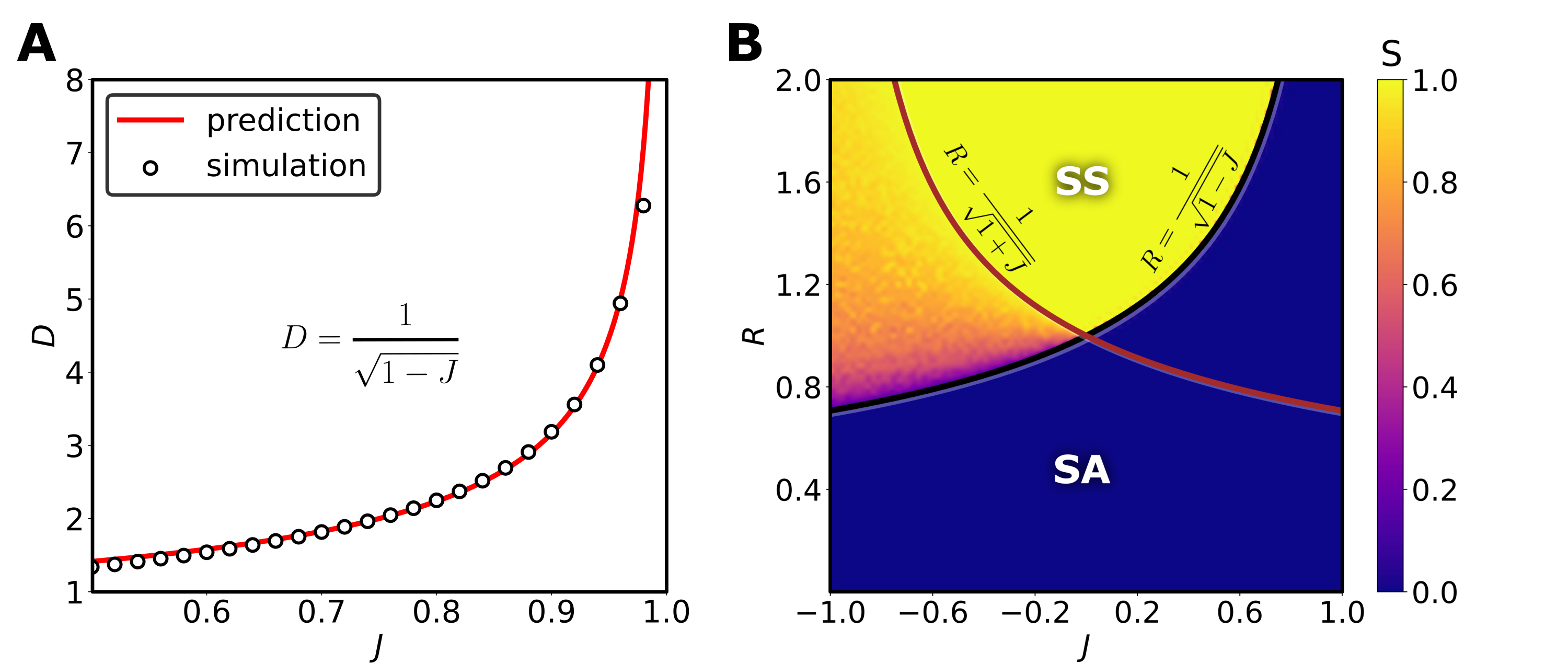}
    \caption{\textbf{Dynamics of two swarmalators.} \textbf{A.} Cluster separation of two-cluster state with varying $J$. The solid red line is the analytical prediction and the circular markers are the simulated results. \textbf{B.} Dynamics of swarmalator model for N=2 with equal attractive and repulsive interactions among the
    orientation vectors.  Heat map illustrates the synchronization order parameter estimated from the evolution of the orientation vectors. The sufficient condition 
    for synchronization was shown to be $R>\vert\bm{x_{12}}\vert=1/{\sqrt{1+J(\bm{\sigma_1}\cdot\bm{\sigma_2})}}$ in  Sec.~S6. The analytical critical curve
    $R=1/{\sqrt{1\pm J}}$, depicted in the figure, matches fairly with the simulation results.}
\end{figure}

\pagebreak
\begin{figure}
    \centering
    \includegraphics[width=\textwidth]{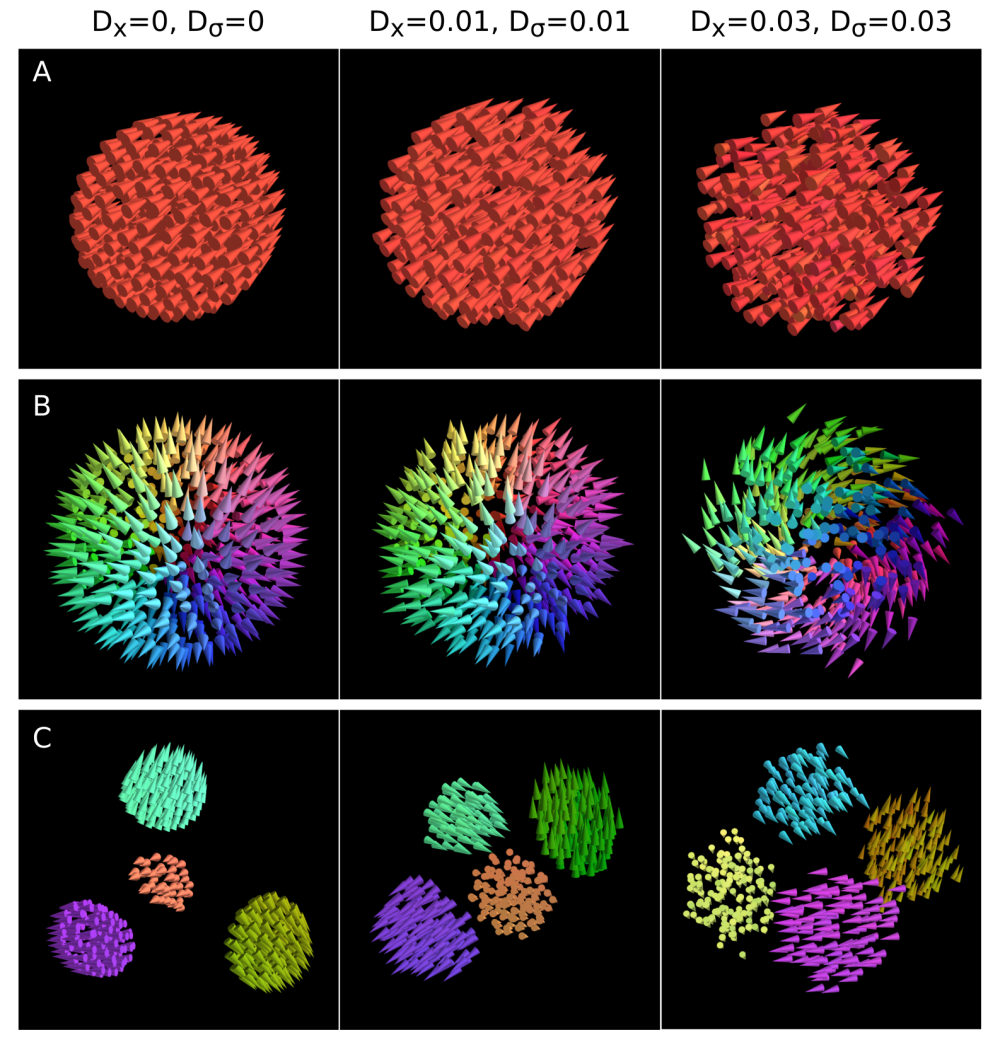}
   \caption{\textbf{Effect of noise on the dynamical states.}  The figure demonstrates the effect of the  white noise for increasing noise strength ($D_x=D_{\sigma}=0.00, 0.01 \text{ and } 0.03$). Note that here same value for the noise strength was used for all $x,y,z$ components of $D_{\bm{x}}$ and $D_{\bm{\sigma}}$. (\textbf{A}) Static sync state.  (\textbf{B})  Spiky state.  (\textbf{C})  Cluster state.
  It is evident that all these three states are  quite robust against the noise. Note that all the other reported collective dynamical states are also found to
  be robust against the employed degree of the noise.}
  \label{gaussian}
\end{figure}

\pagebreak
\begin{figure}[h]
    \centering
    \includegraphics[height=0.8\textheight]{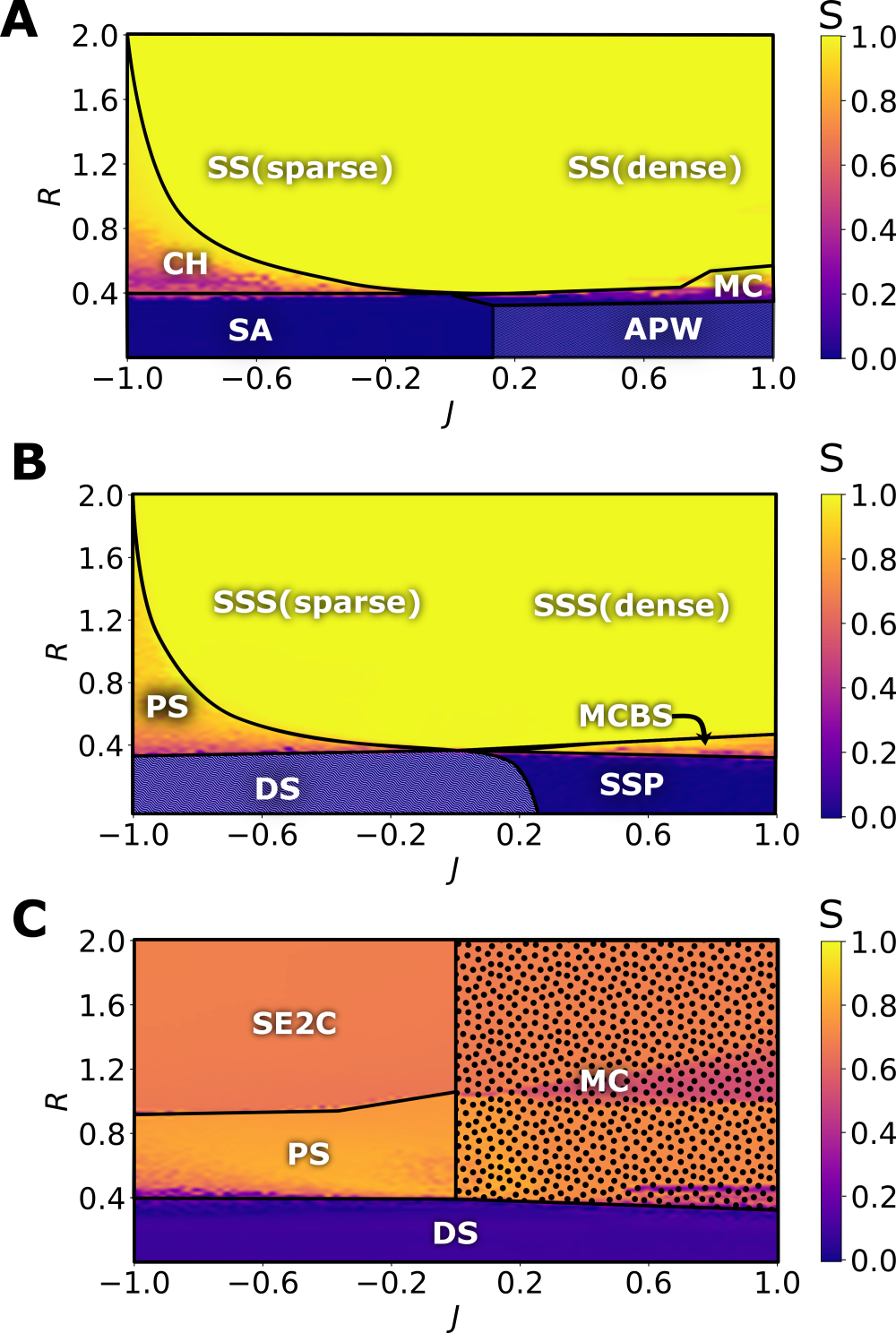}\\
    \caption{\textbf{Attraction dominated competitive  interaction.} Phase diagram in the $(J, R)$ parameter space for  $N=100$ swarmalator collectives when 
$\varepsilon_a=0.9$ and $\varepsilon_r=0.1$
            (\textbf{A}) In the absence of angular frequency  $\bm{\omega}=0$. 
    (\textbf{B})  Orthogonal angular frequencies $\bm{\omega}_1\perp \bm{\omega}_2$.
    (\textbf{C}) Distributed  orthogonal angular frequencies $\bm{\omega}_1\perp \bm{\omega}_2$.
    The self-organizing emergent states observed in the depicted phase diagrams are static async (SA), active phase wave (APW),   chimera (CH), multi-cluster (MC),  synchronized states (SS), disordered spin (DS), spinning spiky states (SSP), pumping state (PS), synchronized spinning state (SSS),  multi-cluster bouncing state (MCBS) and  static embedded two-cluster (SE2C).}
\label{adcpi}
\end{figure}

\pagebreak
\begin{figure}[h]
    \centering
    \includegraphics[height=0.75\textheight]{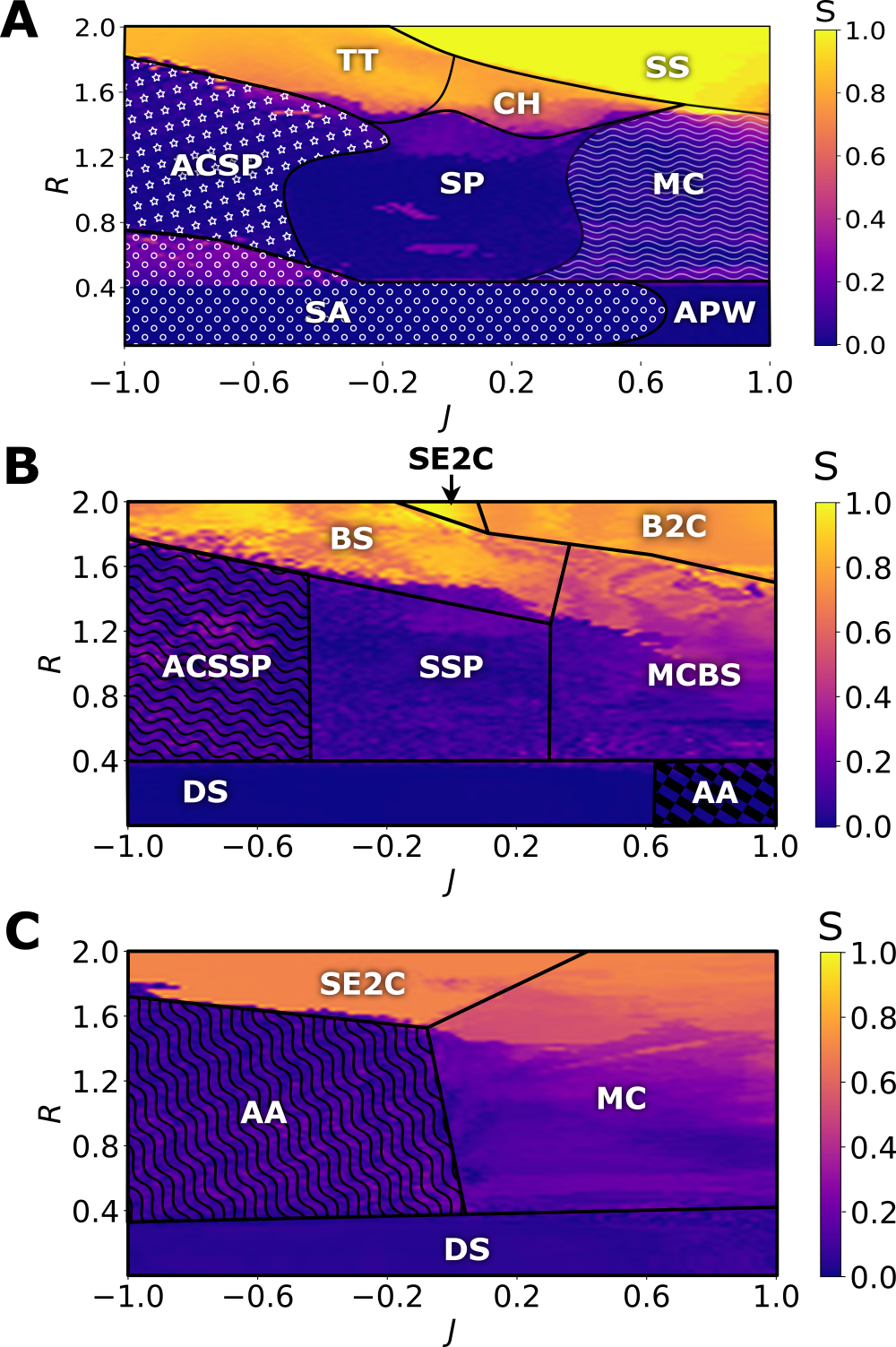}\\
    \caption{\textbf{Competitive  interaction with equal attractive and repulsive coupling strengths.}
    Phase diagram in the $(J, R)$ parameter space for  $N=100$ swarmalator collectives when $\varepsilon_a=\varepsilon_r=0.5$.
            (\textbf{A}) In the absence of angular frequency  $\bm{\omega}=0$. 
    (\textbf{B})  Orthogonal angular frequencies $\bm{\omega}_1\perp \bm{\omega}_2$.
    (\textbf{C}) Distributed  orthogonal angular frequencies $\bm{\omega}_1\perp \bm{\omega}_2$.
    The self-organizing emergent states observed in the depicted phase diagrams  are   static async (SA), active phase wave (APW),  multi-cluster (MC),  spiky state (SP), active core static spiky state (ACSPW), turning tube (TT), chimera (CH), synchronized states (SS), disordered spin (DS), active async (AA), bouncing multi-cluster (BMC), spinning spiky states (SSP), 
 active core spinning spiky state (ACSSP),  breathing state (BS), bouncing two-cluster state (B2C),     
 and  static embedded two-cluster (Static E2C).}
   \label{cpweaer} 
    \end{figure}

\pagebreak
\begin{figure}[h]
    \centering
    \includegraphics[height=0.75\textheight]{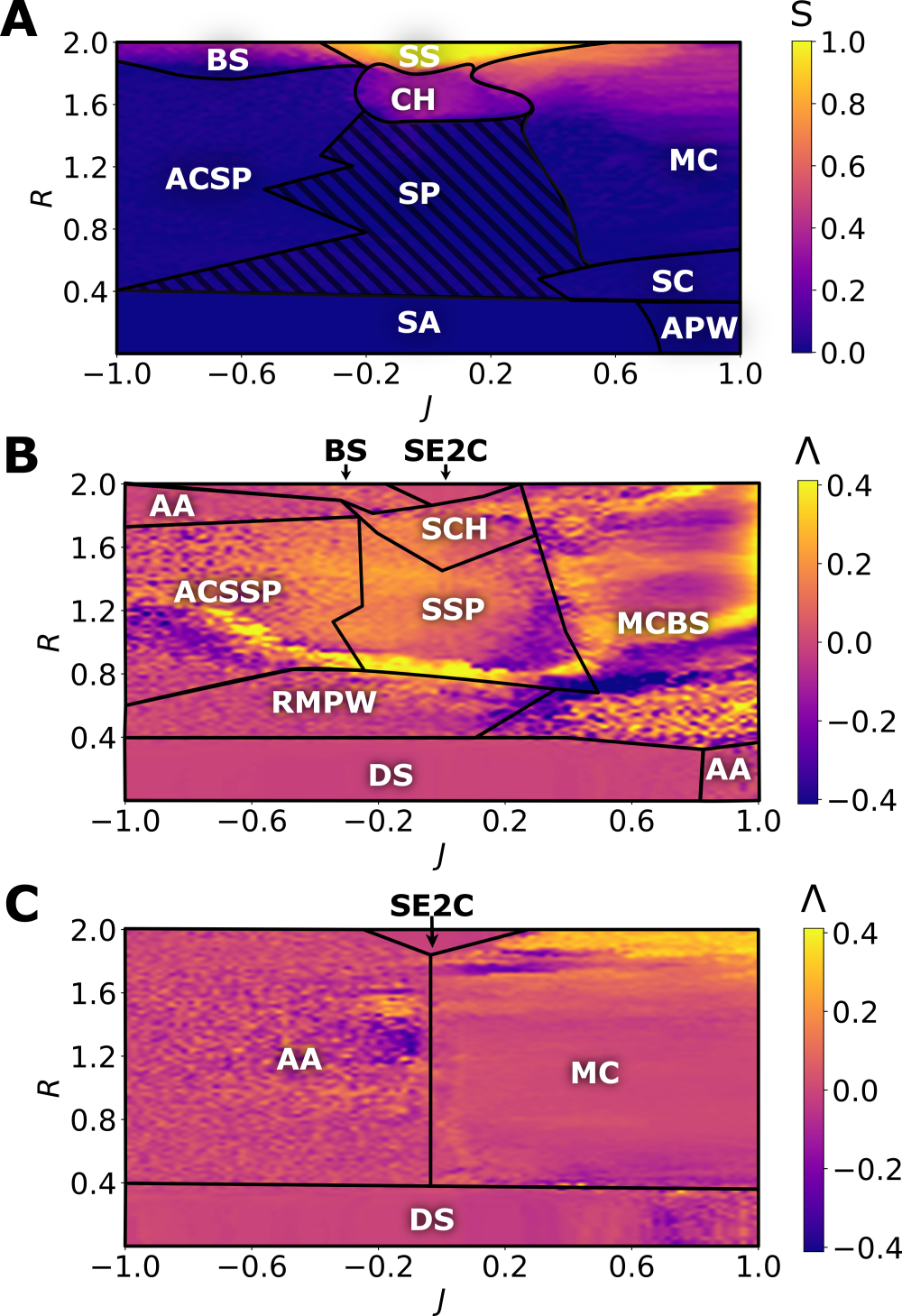}\\
    \caption{\textbf{Repulsion dominated competitive  interaction.} Phase diagram in the $(J, R)$ parameter space for  $N=100$ swarmalator collectives when 
$\varepsilon_a=0.1$ and $\varepsilon_r=0.9$
            (\textbf{A}) In the absence of angular frequency  $\bm{\omega}=0$. 
    (\textbf{B})  Orthogonal angular frequencies $\bm{\omega}_1\perp \bm{\omega}_2$.
    (\textbf{C}) Distributed  orthogonal angular frequencies $\bm{\omega}_1\perp \bm{\omega}_2$.
    The self-organizing emergent states observed in the depicted phase diagrams  are   static async (SA), active phase wave (APW),  spinning cluster (SC), multi-cluster (MC),  spiky state (SP), active core static spiky state (ACSPW),  breathing state (BS), chimera (CH), synchronized states (SS), disordered spin (DS), active async (AA), bouncing multi-cluster (BMC), spinning spiky states (SSP),  active core spinning spiky state (ACSSP),  static two-cluster (S2CS), spinning chimera (SCH), attractive mixed phase wave (AMPW),  and  static embedded two-cluster (Static E2C).}
\label{rdcpi}
\end{figure}

\pagebreak
 \begin{figure}[h]
     \centering
     \includegraphics[width=1\textwidth]{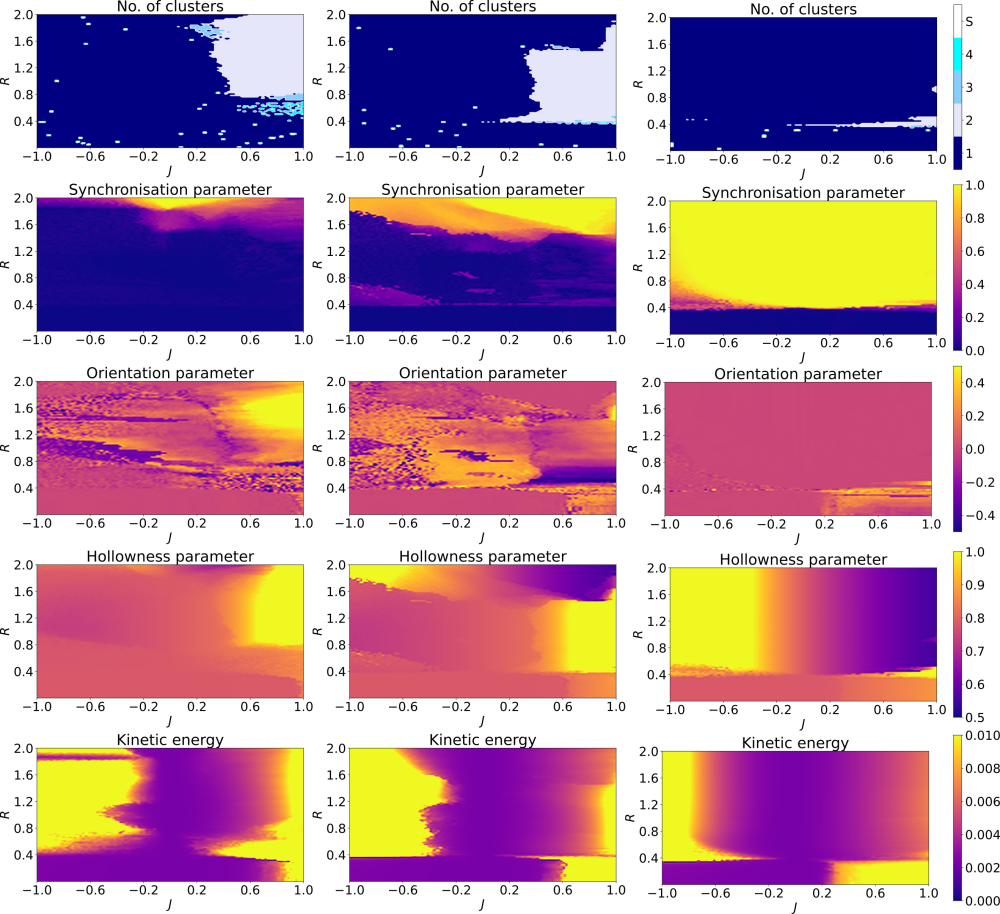}
     \caption{\textbf{Heat maps of  the order parameters for competitive interaction.}  Heat maps in $(J, R)$ parameter space for $N=100$ swarmalator collectives for $\varepsilon_a=\varepsilon_r=0.5$  of  all the order parameters used to characterize distinct dynamical states in the phase diagrams.}
 \end{figure}

\pagebreak
\begin{figure}[h]
    \centering
    \includegraphics[height=0.8\textheight]{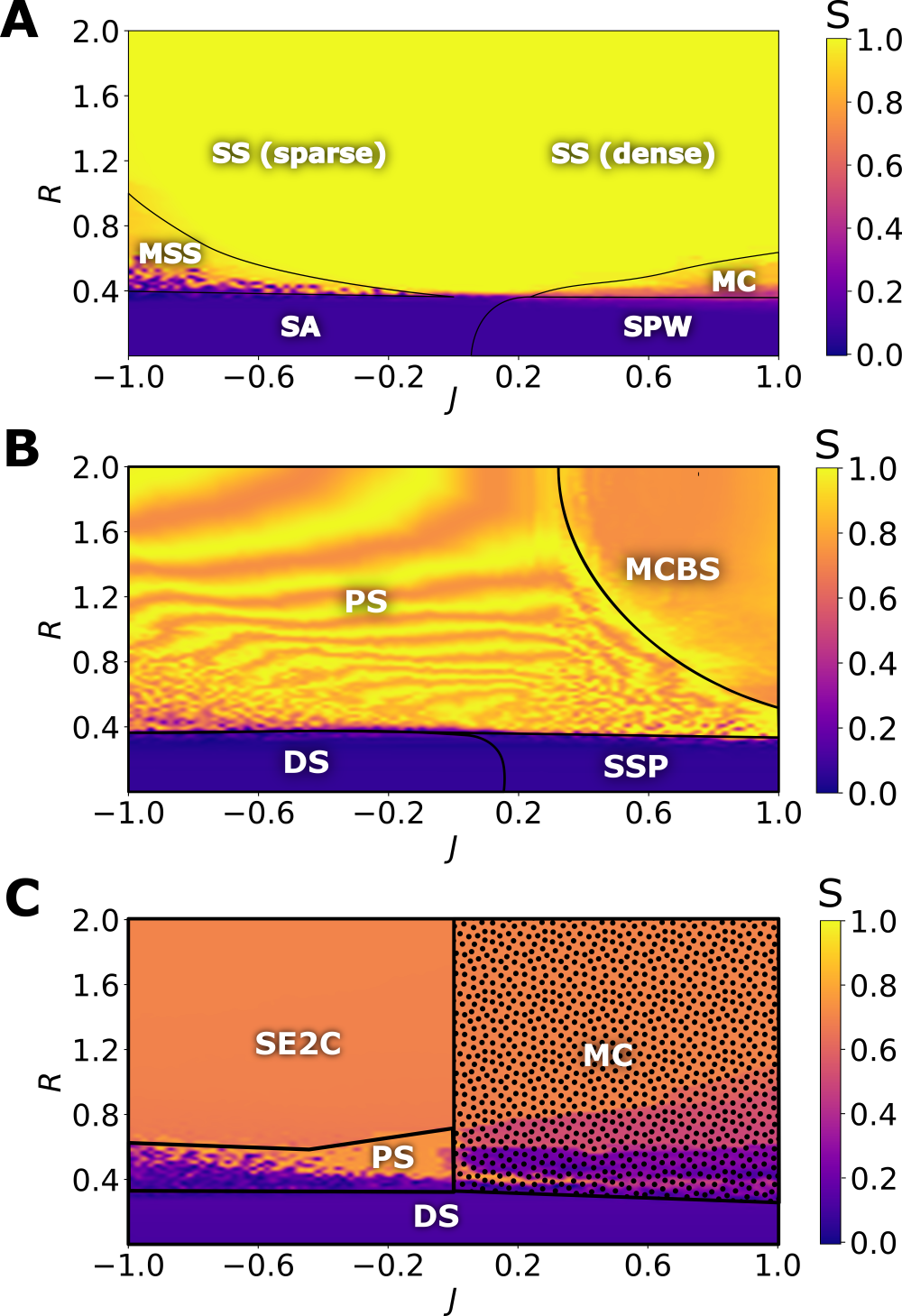}\\
    \caption{\textbf{Purely local attraction.}  Phase diagram in the $(J, R)$ parameter space for  $N=100$ swarmalator collectives when $\varepsilon_r=0$
    and $\varepsilon_a=0.5$.
            (\textbf{A}) In the absence of angular frequency  $\bm{\omega}=0$. 
    (\textbf{B})  Orthogonal angular frequencies $\bm{\omega}_1\perp \bm{\omega}_2$.
    (\textbf{C}) Distributed  orthogonal angular frequencies $\bm{\omega}_1\perp \bm{\omega}_2$.
   The self-organizing emergent states observed in the depicted phase diagrams  are   static async (SA), static phase wave (SPW),  mixed synchronized state (MSS), multi-cluster (MC), sparse and dense synchronized states (SS), disordered spin (DS), spinning spiky states (SSP), pumping state (PS),  multi-cluster bouncing state (MCBS) and  static embedded two-cluster (Static E2C).}
    \label{pli}
\end{figure}

\pagebreak
\begin{figure}[h]
    \centering
    \includegraphics[height=0.8\textheight]{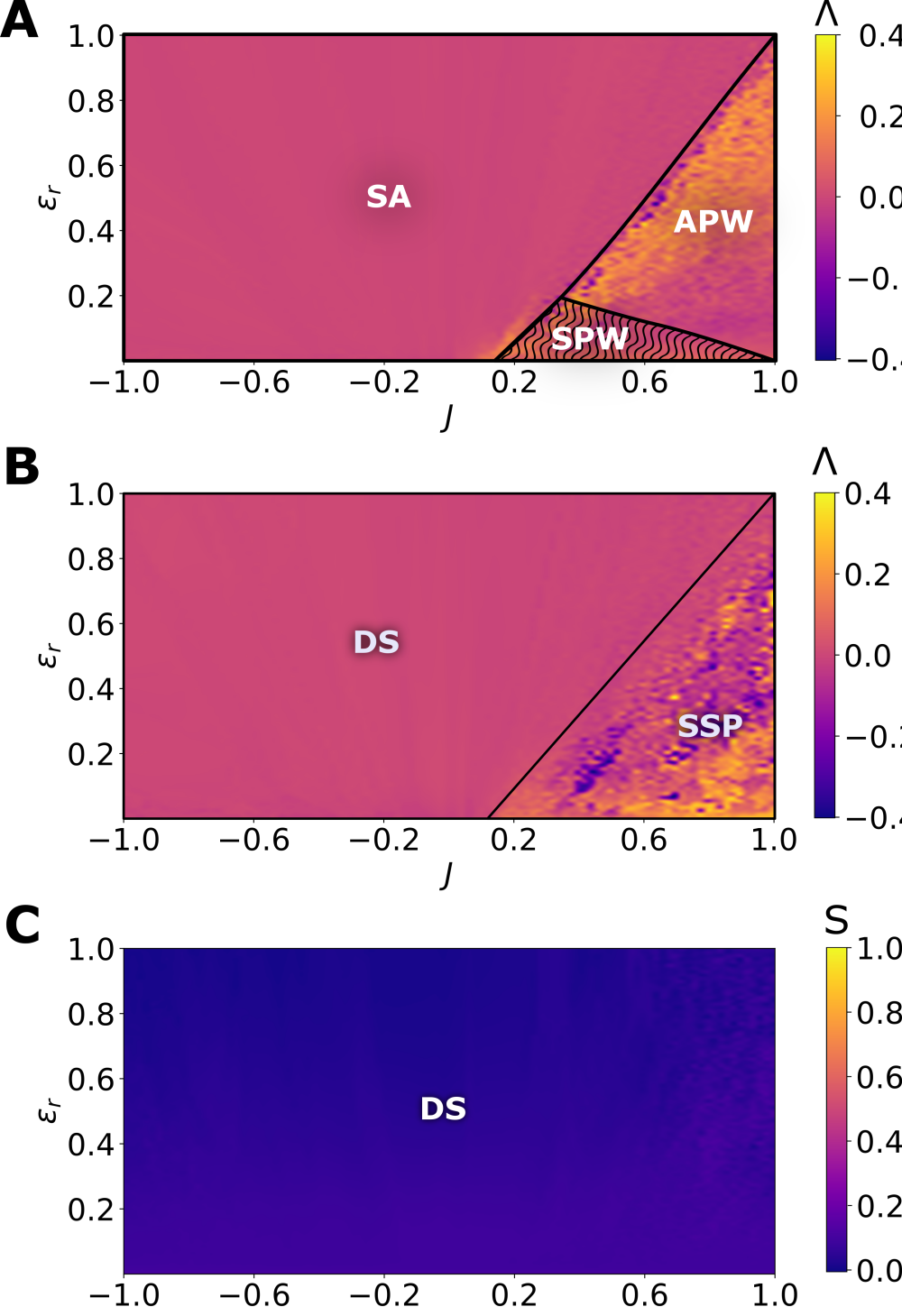}\\
    \caption {\textbf{Global repulsion interaction.}  Phase diagram in the $(J, \varepsilon_r)$ parameter space for  $N=100$ swarmalator collectives and  $R=0.0$.
            (\textbf{A}) In the absence of angular frequency  $\bm{\omega}=0$. 
    (\textbf{B})  Orthogonal angular frequencies $\bm{\omega}_1\perp \bm{\omega}_2$.
    (\textbf{C}) Distributed  orthogonal angular frequencies $\bm{\omega}_1\perp \bm{\omega}_2$.
           The self-organizing emergent states observed in the depicted phase diagrams  are   static async (SA), static phase wave (SPW), active phase wave (APW), disordered spin (DS), and  spinning spiky states (SSP).}
    \label{gri}
\end{figure}

\pagebreak
\begin{figure}
    \centering
    \includegraphics[height=0.8\textheight]{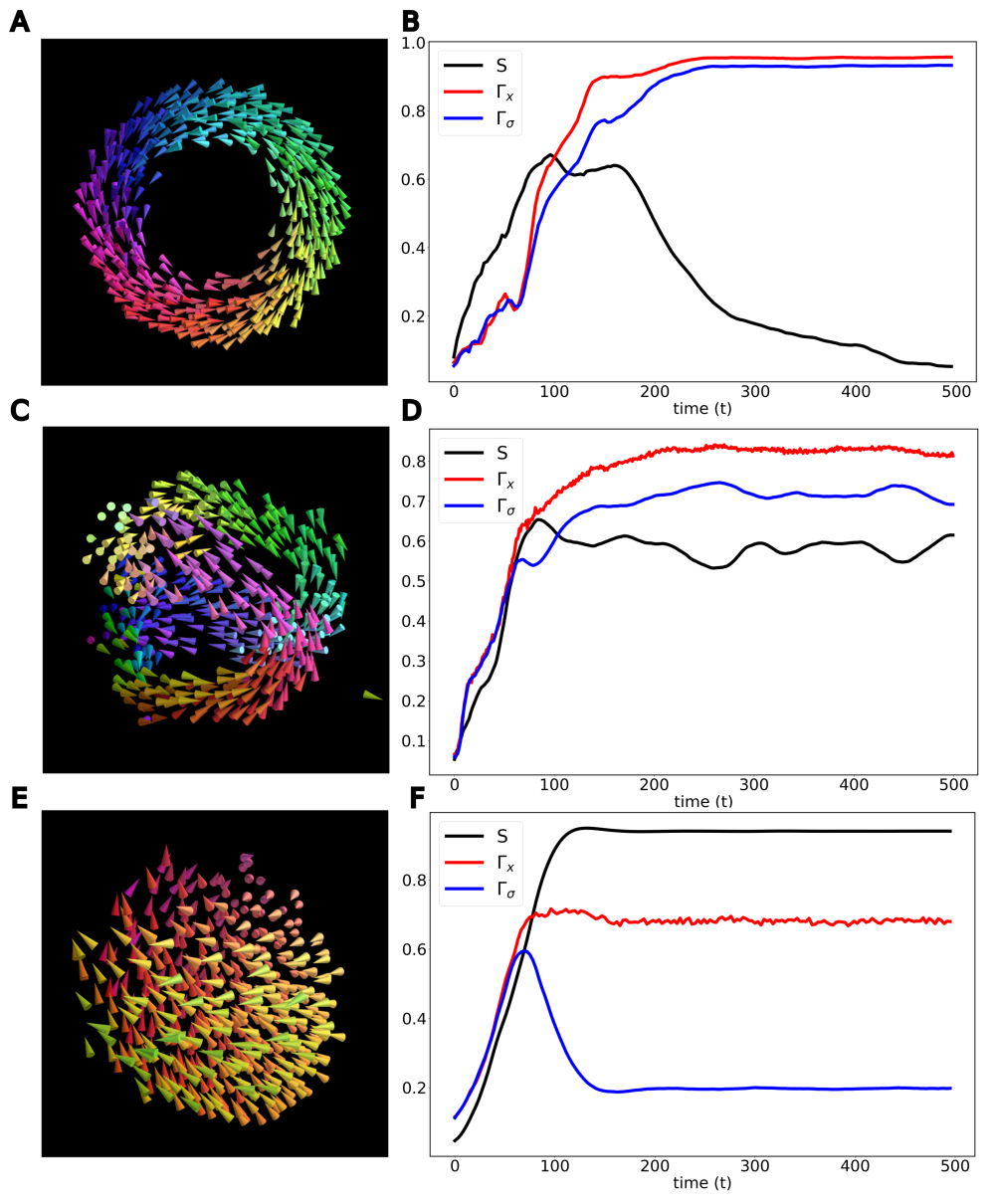}
    \caption{\textbf{Schooling of fish.} 
    Simulations are performed for $N=500$ with the Gaussian noise with noise strength ($D_{x_k} =0.005, D_{\sigma_k}=0.005$). The corresponding plots on the right show the evolution of macroscopic parameters $(S,\Gamma_x,\Gamma_{\sigma})$ with time characterizing the collective behavior of the school. 
    ($c=0.5, \text{ box size} = 1.5,\Delta=0.8,\text{ bound size} =\sqrt3/2 \text{*box size}+\Delta)$.
    (\textbf{A} and \textbf{B}) Milling behavior is observed for $R=0.2, J=0.5$, $\varepsilon_a=0.9, \varepsilon_r=0.0$.
    Simulations are performed for $N=500.$ (\textbf{C} and \textbf{D}) Stripes formation for $R=0.2, J=0.3, \varepsilon_a=0.5, \varepsilon_r=0.0$. (\textbf{E} and \textbf{F}) Rotating crystal for $N=500, R=1.0,$ $ J=0.9, \varepsilon_a=0.3, \varepsilon_r=0.0$. Refer supplementary text~S11 for discussions about the model and the order parameters.}
\end{figure}

\pagebreak
\begin{figure}
    \centering
    \includegraphics[width=\textwidth]{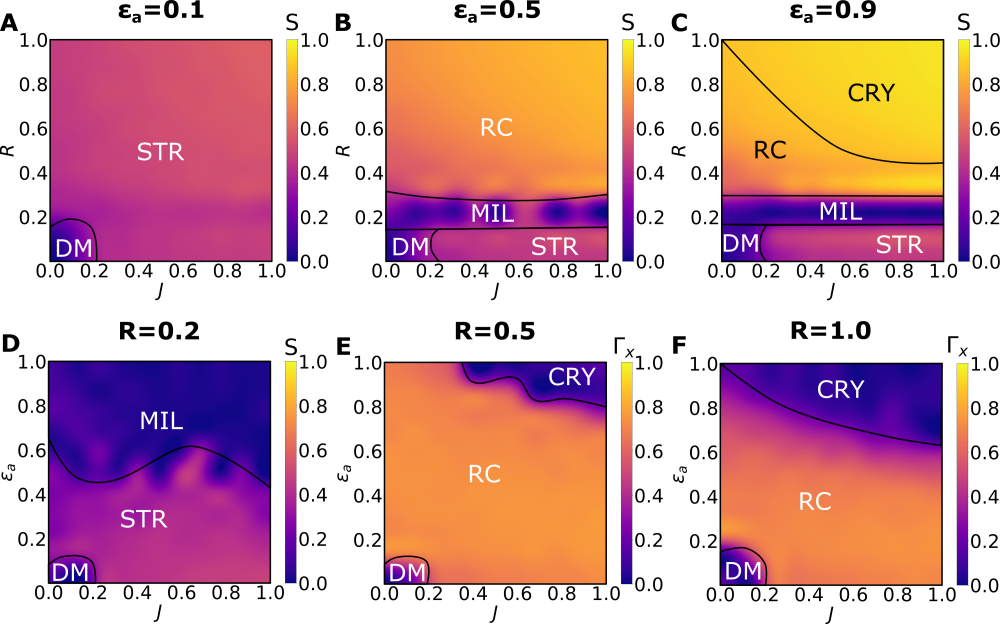}
    \caption{\textbf{Phase diagrams for school of fish.}  Two parameter phase diagrams of school of fish for  $N=100$  (\textbf{A}) in $(J, R)$ parameter space and (\textbf{B}) in $(J, \varepsilon_a)$ parameter space.  The self-organizing emergent states observed in the depicted phase diagrams  are stripes (STR), 
    milling (MIL), rotating crystal (RC), and  disorder motion (DM).}
\end{figure}

\pagebreak
\begin{figure}
    \centering
    \includegraphics[width=\textwidth]{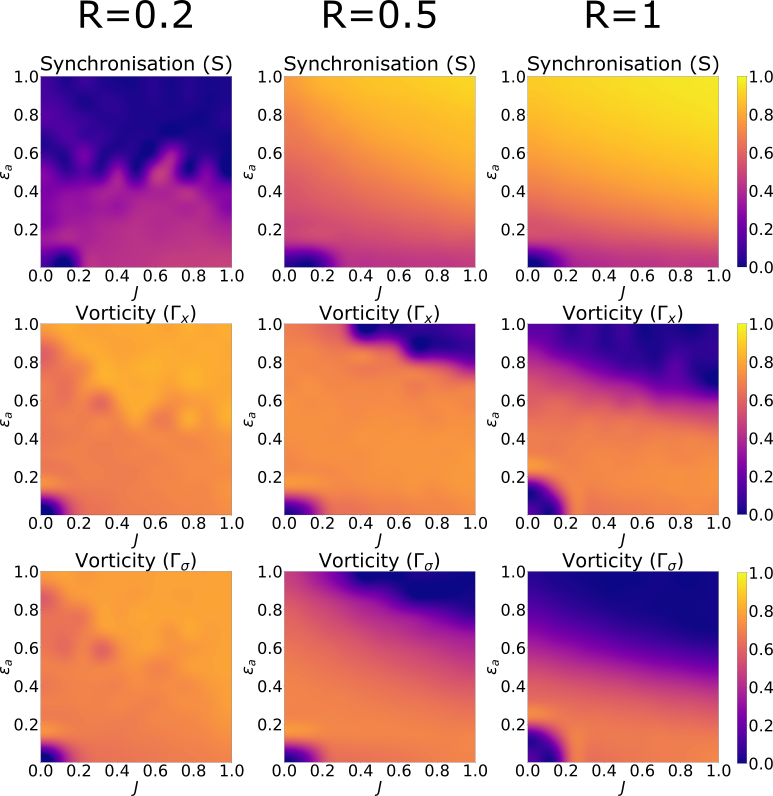}
    \caption{\textbf{Heat maps for school of fish.}  Heat maps of the synchronization order parameter ($S$), spatial vorticity parameter ($\Gamma_x$) and phase vorticity ($\Gamma_\sigma$) in $(J, \varepsilon_a)$ parameter space.}
\end{figure}

\pagebreak
\begin{figure*}[h]
    \centering
    \includegraphics[width=0.9\textwidth]{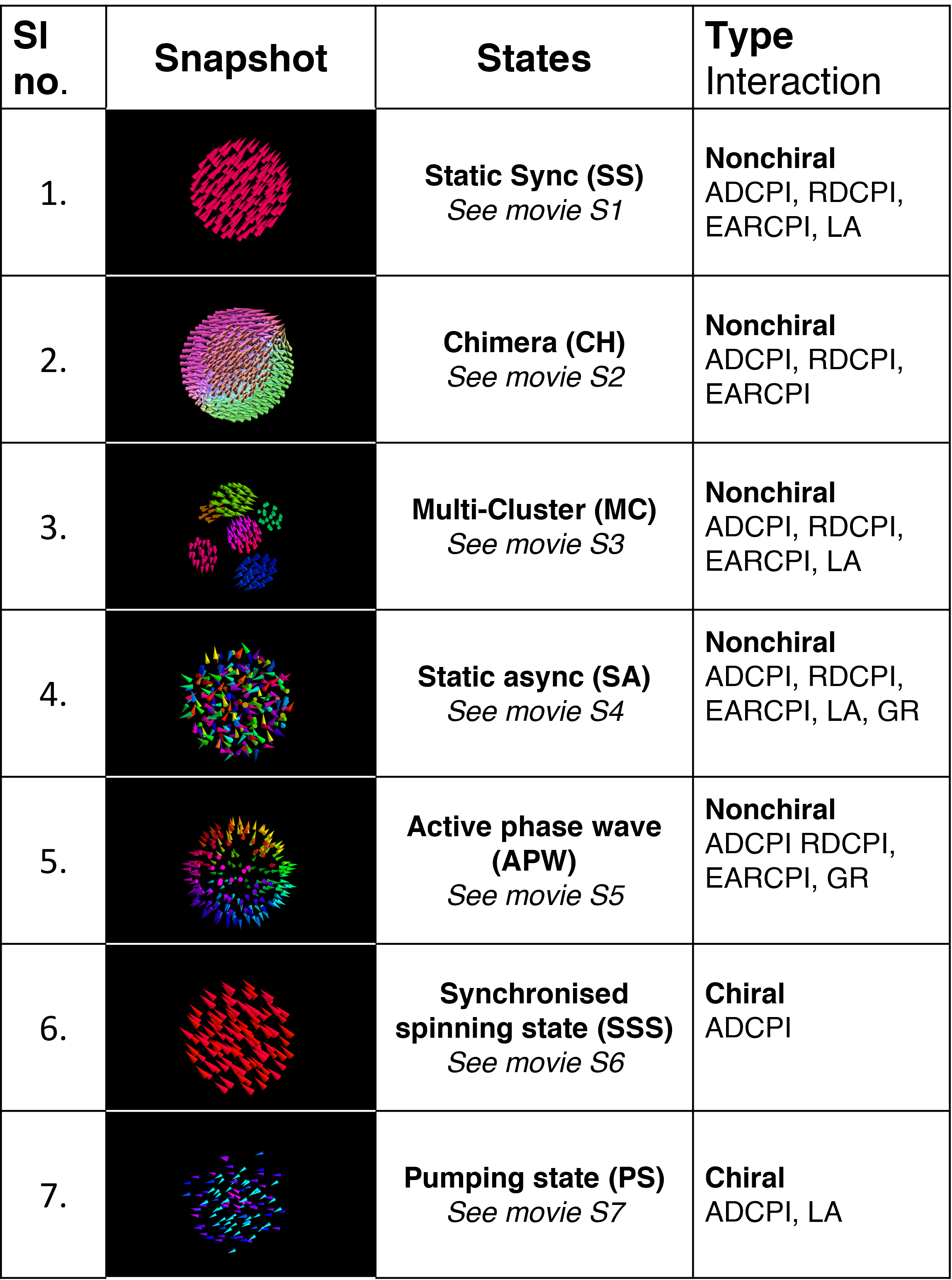}
\end{figure*}

\pagebreak
\begin{figure*}[h]
    \centering
    \includegraphics[width=0.9\textwidth]{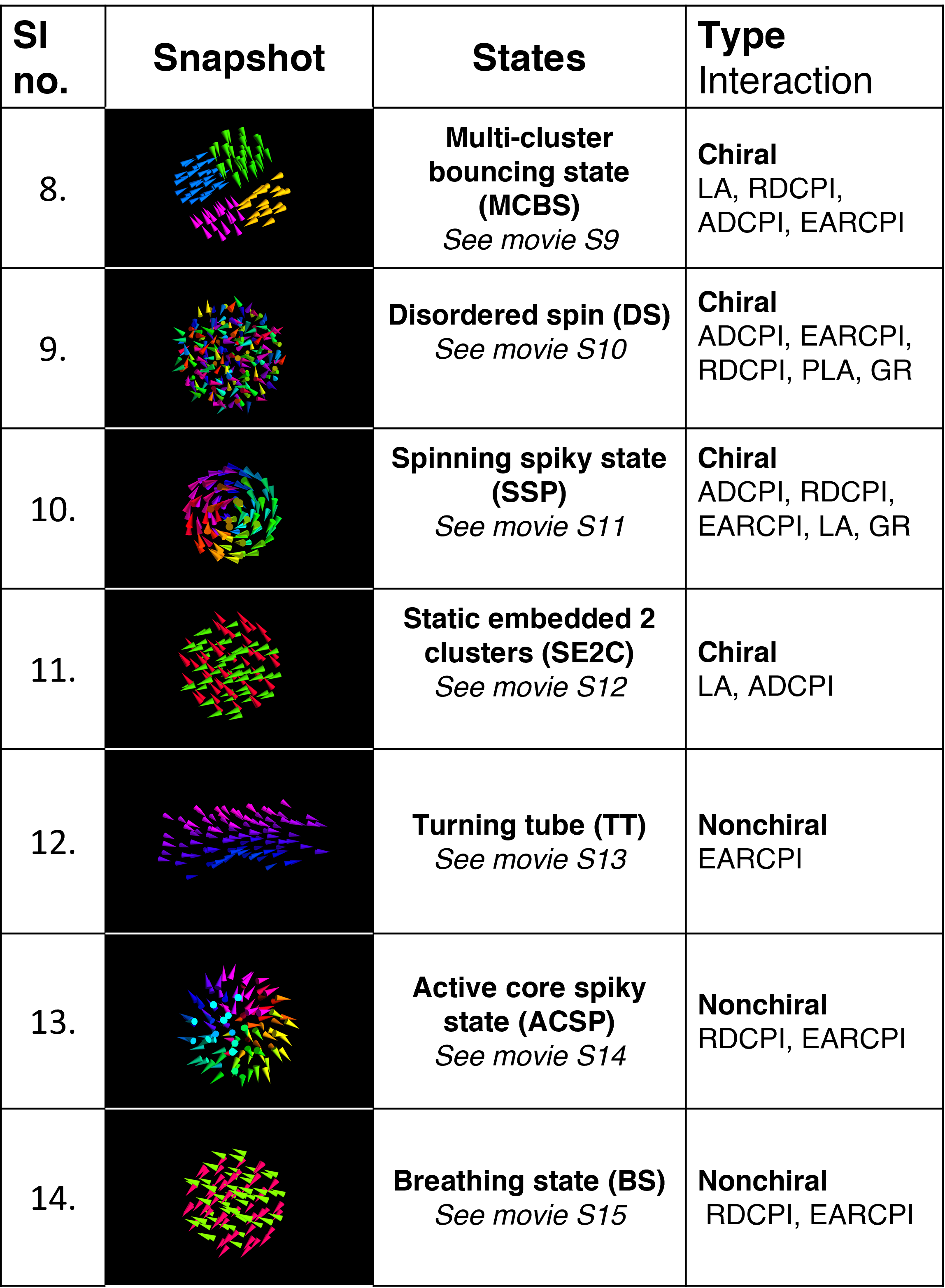}
\end{figure*}

\pagebreak
\begin{figure*}[h]
    \centering
    \includegraphics[width=0.9\textwidth]{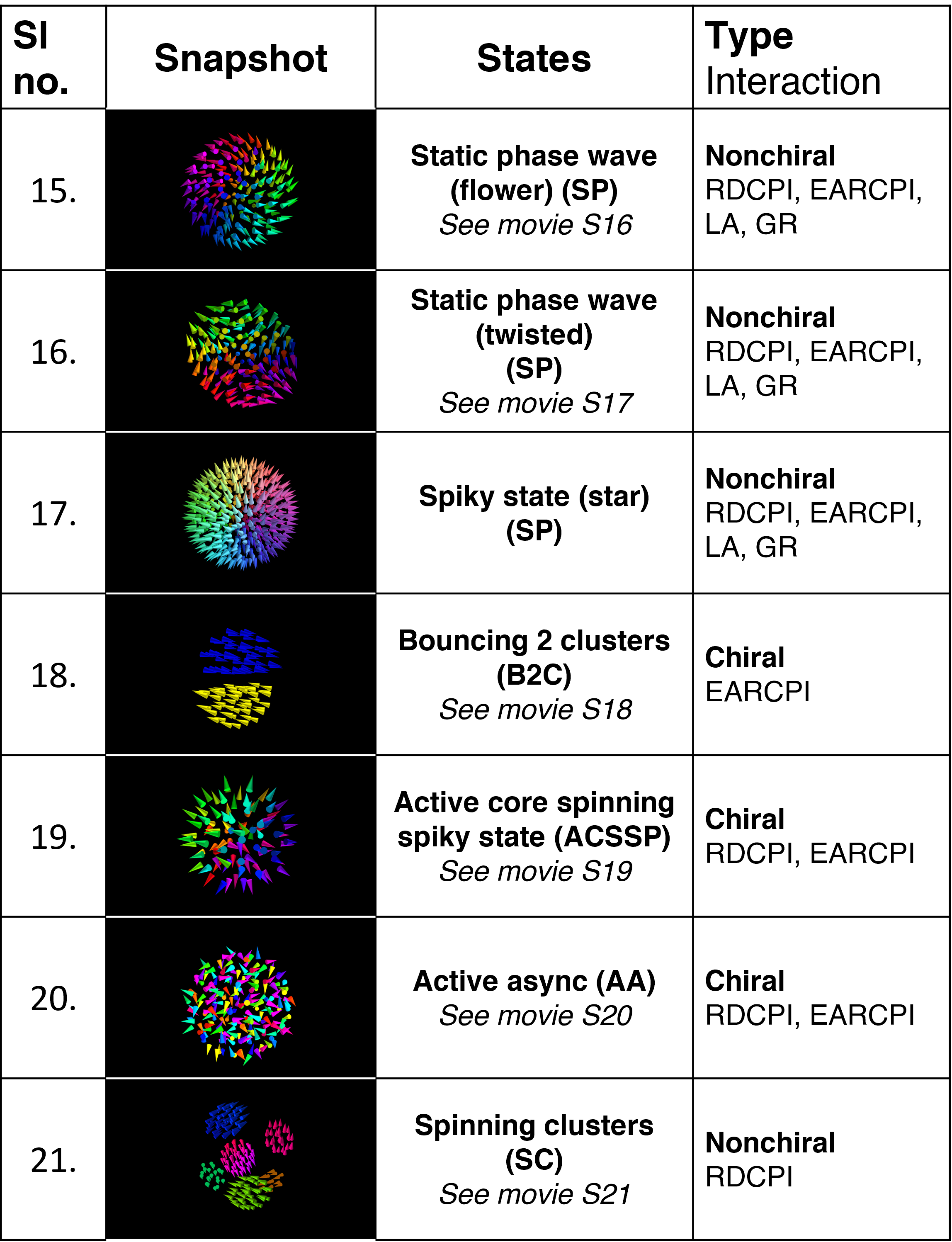}
\end{figure*}

\pagebreak
\begin{figure*}[h]
    \centering
    \includegraphics[width=0.9\textwidth]{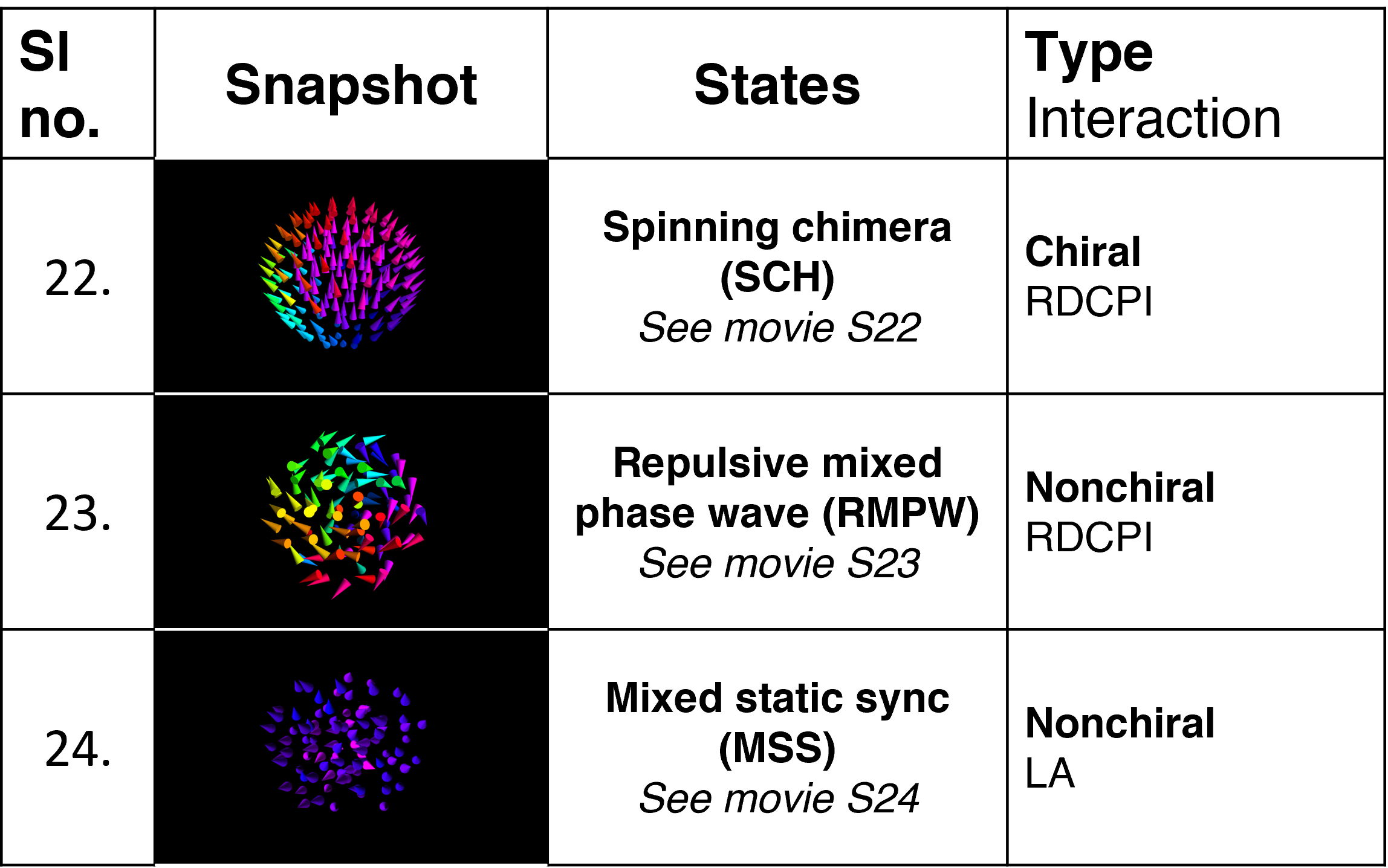}
   
    \parbox{\textwidth}{
        \vspace{5pt}
        \textbf{Table~S1.} \textbf{Gallery of observed self-organizing collective states.} A zoo of self-organizing collective states observed in the vast range of parameters are listed in the table. `Nonchiral' corresponds to the interactions among the swarmalators in the absence of angular frequency for the orientation vectors, whereas `chiral' corresponds to the interactions among the swarmalators with orthogonal angular frequencies. Furthermore, `ADCPI' refers to `attractive dominated competitive phase interaction', `RDCPI' refers to `repulsion dominated competitive phase interaction', `EARCPI' refers to `equal repulsion and attraction competitive phase interaction', `LA' refers to `local attraction', and `GR' refers to `global repulsion'.}
\end{figure*}

\clearpage
\section*{\href{https://drive.google.com/file/d/1bSqgRuZqAMhypK7yYmKIo3C7D5bqLoQA/view?usp=share_link}{Movie S1.}}
\textbf{Static sync.} The movie shows the time evolution of static sync state (SS). Initially randomly distributed swarmalators synchronize their orientations and congregate into a single static cluster.

\section*{\href{https://drive.google.com/file/d/1ORjHeeMltHRhLFVMEo43sPZcdY8MiKhn/view?usp=share_link}{Movie S2.}}
\textbf{Chimera state.} The movie shows a complete and cross-sectional view of chimera state (CH) development during the transition from static async to static sync and back as a function of the vision radius $R$. The synchronized core emerges from the center of the static async state and turns into static sync state for higher values of $R$. 

\section*{\href{https://drive.google.com/file/d/1vgFTdrqkH2Ur3J03L43nhxvPXlIjXz-A/view?usp=share_link}{Movie S3.}}
\textbf{Multi-Cluster state.} The movie shows the time evolution of multi-cluster state (MC).

\section*{\href{https://drive.google.com/file/d/1t8O5Rtf3MS5-WSoNfuAyyQp34ROMiglr/view?usp=share_link}{Movie S4.}}
\textbf{Static async.} The movie shows the time evolution of static async state (SA). Initially randomly distributed swarmalators desynchronize their orientations and congregate into a single static cluster.

\section*{\href{https://drive.google.com/file/d/1ChMLh5iMn38h7w4hCXbaR4DHHmsJUC9X/view?usp=share_link}{Movie S5.}}
\textbf{Active phase wave.} The movie shows the time evolution of active phase wave (APW).

\section*{\href{https://drive.google.com/file/d/1mq1Is5MKrrhtRwkub3v0PmmC6mctoqbT/view?usp=share_link}{Movie S6.}}
\textbf{Sync spinning state.} The movie shows the time evolution of synchronized spinning state (SSS). Initially randomly distributed swarmalators synchronize their orientations and congregate into a single spinning (precessing) cluster. This precession effect happens in the presence of orthogonal angular frequencies applied to the population.

\section*{\href{https://drive.google.com/file/d/1bG9PKQYREnrUG26nzQqLIVGNm_1eLIdD/view?usp=share_link}{Movie S7.}}
\textbf{Pumping State.} The movie shows the time evolution of pumping state (PS) observed in the case of two orthogonal angular frequencies. This state involves period-2 oscillation of a single cluster.

\section*{\href{https://drive.google.com/file/d/1S_sLSFBqoSObyFb_CE-L3ckFzMlwzKDH/view?usp=share_link}{Movie S8.}}
\textbf{Oscillation Death.} The movie shows the oscillation death in swarmalators with distributed frequencies.

\section*{\href{https://drive.google.com/file/d/1MPbKSArRI29UgohNhNv5y6p88QQnG6ea/view?usp=share_link}{Movie S9.}}
\textbf{Multi-Cluster bouncing state.} The movie shows the time evolution of the multi-cluster bouncing state (MCBS). The inter-cluster separation increases and decreases periodically. 

\section*{\href{https://drive.google.com/file/d/1iqHkXlXfR9Tx5inMM3VWyBZiHlJwt1rK/view?usp=share_link}{Movie S10.}}
\textbf{Disordered spin state.} The movie shows the time evolution of disordered spin state (DS). The initially randomly distributed swarmalators desynchronize their orientations and congregate into single spinning (precessing) cluster. This precession effect can be attributed to the presence of angular frequencies.

\section*{\href{https://drive.google.com/file/d/1vhx3A2XiAnYK-I5qvRSFTERUCQVqdcpj/view?usp=share_link}{Movie S11.}}
\textbf{Spinning spiky state.} The movie shows the time evolution of spinning spiky state (SSP). This precession effect can be attributed to the presence of two orthogonal angular frequencies.

\section*{\href{https://drive.google.com/file/d/1-DQj3QbeEOwyd_xg828FeRga55awIk8q/view?usp=share_link}{Movie S12.}}
\textbf{SE2C state.} The movie shows the time evolution of static embedded two-cluster state (SE2C). The oscillation death in pumping state results in static embedded two-cluster.

\section*{\href{https://drive.google.com/file/d/1kSemwkiwdisCKv5TmuSsFlLdo_XSN7qX/view?usp=share_link}{Movie S13.}}
\textbf{Turning tube state.} The movie shows the time evolution of turning tube (TT) state. The elongation of static sync state in $J<0$ region with finite vision radius results in two cylindrical formation that shows collective rotation.

\section*{\href{https://drive.google.com/file/d/1AJV3joaRmXiqCima85svmnU93iYqdoIu/view?usp=share_link}{Movie S14.}}
\textbf{Active core spiky state.} The movie shows the time evolution of active core spiky (ACSP) state. The turbulent core emerges from center of static spiky state and turns into active async state for higher values of $R$. 

\section*{\href{https://drive.google.com/file/d/1YEoGgDoJCfyS2EIbuSmnYolqnjvIVLyR/view?usp=share_link}{Movie S15.}}
\textbf{Breathing state.} The movie shows the time evolution of breathing state (BS). The dynamics consists of cyclic expansion and contraction of static embedded two cluster state. 

\section*{\href{https://drive.google.com/file/d/1OLpp0eAfVz3pmTGi0nfcY3BwH34eeeuV/view?usp=share_link}{Movie S16.}}
\textbf{Flower state.} The movie shows the time evolution of flower state (SP).

\section*{\href{https://drive.google.com/file/d/1LRuc4eVtbUUCO3NiT49BKsGtrrHuU3LO/view?usp=share_link}{Movie S17.}}
\textbf{Twisted state.} The movie shows the time evolution of twisted state (SP).

\section*{\href{https://drive.google.com/file/d/1axrK4yhcAPUWIBaHajVhimOe_Z-tEcfM/view?usp=share_link}{Movie S18.}}
\textbf{Bouncing two cluster.} The movie shows the time evolution of bouncing two-cluster (B2C) state.

\section*{\href{https://drive.google.com/file/d/1xTyy5BINqKAskldWA3Jl5BttYteZ-B5w/view?usp=share_link}{Movie S19.}}
\textbf{Active core spinning spiky state.} The movies shows the time evolution of active core spinning spiky (ACSSP) state.  The turbulent core emerges from center of spinning spiky state and turns into an active async state for higher values of $R$. 

\section*{\href{https://drive.google.com/file/d/1GDFfQTqQHrFfIGBOjebPY8Ra-WlnNYIE/view?usp=share_link}{Movie S20.}}
\textbf{Active async.} The movie shows the time evolution of active async state.

\section*{\href{https://drive.google.com/file/d/1Lm_nZRIicj_GtgEzBgMygaS0lHAvKDY3/view?usp=share_link}{Movie S21.}}
\textbf{Spinning cluster state.} The movie shows the time evolution of spinning (rotating) cluster state.

\section*{\href{https://drive.google.com/file/d/1d5xXnetF1c9mDzjVtOfnEElHGbBuGo2k/view?usp=share_link}{Movie S22.}}
\textbf{Spinning chimera state.}  The movie shows the time evolution of spinning chimera state (SCH). A synchronized spinning (precessing) core emerges from the center of the spinning spiky state and turns into synchronized spiky state.

\section*{\href{https://drive.google.com/file/d/1klwwZQoeORZz3lP6bU7rerb4E0A1TDNb/view?usp=share_link}{Movie S23.}}
\textbf{Repulsive Mixed Phase Wave.} The movie shows the time evolution of repulsive mixed phase wave (RMPW). The intermediate vision radius in the  case of repulsive dominated competitive interaction between the orientation vectors   converts the spinning spiky states into turbulent state.

\section*{\href{https://drive.google.com/file/d/1CbtVEkngopVThscAue_TaXsiItZICFO-/view?usp=share_link}{Movie S24.}}
\textbf{Mixed static sync.} The movie shows the time evolution of mixed static sync (MSS). The state consists of synchronized and partially synchronized population. 

\section*{\href{https://drive.google.com/file/d/1UxTxTIFWB_plDuaKVaEgCCd0LEGA8tPQ/view?usp=share_link}{Movie S25.}}
\textbf{Fish schooling.} The movie depicts the polarized and milling states in the extended swarmalator model. In polarized (crystal) state the fish synchronize their orientation, while during milling fish move in circular fashion.

\section*{\href{https://drive.google.com/file/d/1onU0-F2Y7KrJPIe5ZHvOEPFAPEGq1T3r/view?usp=share_link}{Movie S26.}}
\textbf{Fish schooling experimental analysis.} This movie shows the visualization of fish schooling experimental data and corresponding order parameters used for characterizing the schooling dynamics.

\section*{\href{https://drive.google.com/file/d/1RZ1kUJkPDBtpVlrUKUWxctZJNa4vCSUV/view?usp=share_link}{Movie S27.}}
\textbf{Traveling waves of genetic expression.} This movie depicts the traveling phase wave in swarmalators in the presence of local spatial interactions, which resembles the traveling wave of gene expression in mouse embryo during embryonic developmemt of vertebra segments. 

\section*{\href{https://drive.google.com/file/d/1S5rOdiOGTUMbb2C2kcwP_IGW-8Gc_YaV/view?usp=share_link}{Movie S28.}}
\textbf{Cell sorting.} This movie depicts the swarmalator analog of the cell sorting process for two and three types of cells. Random distribution of swarmalators spontaneously sort themselves into layers due to differential spatial coupling.

\section*{\href{https://drive.google.com/file/d/1dY_r9CucnzNo1X5MNZqqkRXHdZUw1-yR/view?usp=share_link}{Movie S29.}}
\textbf{Aggregation in slime mold.} This movie depicts the aggregation of swarmalators in the presence of local spatial interactions. The emerging structure resembles the cellular aggregation of slime mold to form a multi-cellular organism.

\pagebreak

\end{document}